\documentclass[fleqn,usenatbib]{mnras}

\usepackage[T1]{fontenc}

\DeclareRobustCommand{\VAN}[3]{#2}
\let\VANthebibliography\thebibliography
\def\thebibliography{\DeclareRobustCommand{\VAN}[3]{##3}\VANthebibliography}

\usepackage{graphicx}	
\usepackage{amssymb}	

\usepackage{amsmath}	
\usepackage[table,xcdraw]{xcolor}
\usepackage{threeparttable}
\usepackage{ulem}
\usepackage{adjustbox}

\usepackage{newtxtext,newtxmath}

\title[X-ray polarimetry of lamp-post reflection]{Spectral and polarization properties of reflected X-ray emission from
black-hole accretion discs for a distant observer: the lamp-post model}

\author[J. Podgorn\'y et al.]{
J. Podgorn\'y$^{1,2,3}$\thanks{E-mail: jakub.podgorny@astro.unistra.fr}, 
M. Dov{\v{c}}iak$^{2}$,
R. Goosmann$^{1}$,
F. Marin$^{1}$,
G. Matt$^{4}$,
A. R\'o\.za\'nska$^{5}$
and V. Karas$^{2}$
\\
$^{1}$Universit\'e de Strasbourg, CNRS, Observatoire Astronomique de Strasbourg, UMR 7550, F-67000 Strasbourg, France\\
$^{2}$Astronomical Institute, Academy of Sciences of the Czech Republic, Bo{\v{c}}n\'i II, CZ-14100 Prague, Czech Republic\\
$^{3}$Astronomical Institute, Charles University, V Hole{\v{s}}ovi{\v{c}}k\'ach 2, CZ-18000 Prague, Czech Republic\\
$^{4}$Dipartimento di Matematica e Fisica, Universit\`a Roma Tre, via della Vasca Navale 84, I-00146 Rome, Italy\\
$^{5}$Nicolaus Copernicus Astronomical Center, Polish Academy of Sciences, Bartycka 18, 00-716 Warsaw, Poland
}

\date{Accepted XXX. Received YYY; in original form ZZZ}

\pubyear{2022}

\begin{document}
\label{firstpage}
\pagerange{\pageref{firstpage}--\pageref{lastpage}}
\maketitle

\begin{abstract}

Rebirth of  X-ray polarimetric instruments will have a significant impact on our knowledge of compact accreting sources. The properties of inner-accreting regions of active galactic nuclei (AGNs) or X-ray binary systems (XRBs), such as black-hole spin, their disc inclination and orientation, shape and size of their corona, can be polarimetrically studied, parallelly to the well-known X-ray spectroscopic and timing techniques. In this work, we provide a new spectropolarimetric numerical estimate of X-rays in the lamp-post coronal model for a distant observer, including a polarized reflected radiation from the accretion disc. The local disc reflection was simulated using the codes {\tt TITAN} and {\tt STOKES} and includes variable disc ionization as well as Monte Carlo treatment of Compton multiple scatterings. We introduce a relativistic code {\tt KYNSTOKES} based on our well-tested {\tt KY} package that accounts for all relativistic effects on radiation near a black hole, apart from the returning radiation, and adds a possibility of polarized coronal emission. We study the spectrum, polarization degree and polarization angle at spatial infinity for various global system parameters and we demonstrate the difference at infinity, if analytical local reflection computations are used. We newly predict that in the hard X-rays the reflected component can be 25\% polarized and the total emission can be 9\% polarized in the most favourable, yet realistic configurations of radio-quiet AGNs. Thus, the relativistic disc reflection remains important for the interpretation of X-ray polarimetric observations.

\end{abstract}




\section{Introduction}

X-ray polarimetric and spectroscopic missions planned for this and the next decade will significantly enhance our knowledge of compact accreting sources. Whether it is the polarimetric missions: IXPE [\cite{Weisskopf2022}, launched successfully on December 9th, 2021] or eXTP [\cite{Zhang2016, Zhang2019}, due to be launched in the second half of 2020s], or, for example, the spectroscopic ATHENA mission [\cite{Barret2020}, due to be launched in 2030s], the new instruments on board of such satellites will significantly expand the current windows into the high-energy Universe. Thus, it is high time to revisit theoretical models of X-ray emission from the two most common classes of X-ray accreting objects: the supermassive black holes (BHs) in active galactic nuclei (AGNs) and the stellar-mass BHs in X-ray binary systems (XRBs) \cite[see e.g.][]{Antonucci1993,Done2007,Trumper2008,Seward2010,Abramowicz2013,Fabiani2014,Netzer2015}.

In particular it is the polarization models of AGNs and XRBs that require attention, because the field of X-ray polarimetry is currently re-opening after being silent for almost 40 years \citep{Fabiani2014}. The IXPE mission has already brought a handful of high-quality X-ray polarimetric observations of AGNs \citep{Marinucci2022, Ursini2023, Gianolli2023, Ingram2023, Tagliacozzo2023} and XRBs \citep{Krawczynski2022c, Veledina2023, Podgorny2023, Rawat2023, Kushwaha2023, Ratheesh2023, Cavero2023}. Studying the X-ray polarization signal will bring an independent method for determining the observer inclination and source orientation on the sky \citep{Dovciak2004, Dovciak2008, Li2009, Dovciak2011, Marin2014, Marin2016} or determining the spin of BHs \citep{Connors1977,Stark1977,Connors1980, Dovciak2008, Schnittman2009, Li2009, Taverna2020, Taverna2020b}, complementary to the well-known spectroscopic (using either the iron K$\alpha$ line profile or the thermal disc continuum emission) and timing (kHz QPOs) techniques [see \cite{Reynolds2019} and references therein].

In our study, we focus on the inner-accretion models of radio-quiet sources. The concept of a hot gaseous corona above the accretion disc has been investigated in the past decades and put forward as a promising explanation for the primary X-ray power-law emission typically observed in AGNs and XRBs \citep[see e.g.][]{Sunyaev1980, Haardt1993, Haardt1993b, Dove1997, Krolik1999, Seward2010}. The Compton up-scattering of thermal disc photons on the coronal electrons produces X-ray signal that can reach directly the observer or that can be further re-processed by the disc or by more distant components of the system. The primary X-rays can already obtain partial scattering-induced polarization ($\lesssim 10 \%$) in the corona \citep{Poutanen1996, Schnittman2010, Marinucci2018, Beheshtipour2017, Tamborra2018, BeheshtipourThesis, Ursini2022, Krawczynski2022a, Krawczynski2022c, Poutanen2023}. Our objective is to focus on the reflection of the X-ray power-law emission from the accretion disc and to discuss spectral and polarization signatures of such radiation for a distant observer, i.e. after integrating the reflected Stokes parameters in the local co-moving frame over the accretion disc, including all the special and general relativistic (GR) effects in the vicinity of a central black hole when the system is viewed as point source from spatial infinity. The most important process that should significantly enhance the polarization degree of the total signal in the hard X-rays is Compton down-scattering of the coronal photons in the disc \citep{Matt1993b, Poutanen1996a, Schnittman2009, Dovciak2011, Podgorny2021}. We may then add the direct primary radiation that reaches a distant observer through the same strong-gravity environment and construct toy models of inner-accreting X-ray emission of AGNs and XRBs.

The integration and ray-tracing processes require choosing some prescribed simple coronal geometry, which is until today one of the main unknowns of such modeling \citep[see e.g.][]{Marinucci2018,Poutanen2018, Kubota2018}. Several global geometries of the corona can be considered. In this paper, we would like to explore the outcome of the most simple coronal archetype: the \textit{lamp-post model} \citep{Matt1991,Martocchia1996,Henri1997,Petrucci1997, Martocchia2000, Miniutti2004, Dovciak2004b, Dovciak2011, Parker2015, Furst2015, Miller2015, Niedzwiecki2016, Dovciak2016, Walton2017, Ursini2020}. The lamp-post model assumes a small patch of a hot corona, sometimes interpreted as a failed jet, located on the rotational axis of the disc at some height above the central black hole. We represent an isotropic point-like source with stationary power-law emission. Although non-stationarity of the primary source is sometimes considered \citep{Beloberodov1999, Malzac2001}, here we assume a stationary situation. Other classes of models assume e.g. the \textit{extended corona} \citep{Haardt1991, Haardt1993, Poutanen1996, Dabrowski2001,Malzac2005,Niedzwiecki2008, Schnittman2010, Marinucci2018, Poutanen2018} (a diffuse optically thin hot corona near above large area of the disc ranging from slab to spherical assumptions) or a truncated inner accretion disc (i.e. not extending to the innermost stable circular orbit) with the so-called \textit{hot inner accretion flow}, which can be divided into sub-classes based on plasma interaction of the hot flow with the inner edge of the cold disc \citep{Esin1997, Esin1998, Poutanen1996, Dove1997, Zdziarski1998, Liu2007, Veledina2013, Poutanen2009, Poutanen2018}.

In the literature, one can find X-ray local spectral reflection models assuming a stratified disc atmosphere in hydrostatic equilibrium accomplished with the codes {\tt TITAN} and {\tt NOAR} \citep{Rozanska2002,Dumont2003}, or with the {\tt ATM24} code \citep{rozanska2008,Rozanska2011,vincent2016}. Although the constant density assumption is still until today a debatable simplification \citep{Nayakshin2000, Nayakshin2001, Pequignot2001, Ballantyne2001, Rozanska2002, Dumont2002, Ross2007, rozanska2008, Rozanska2011, vincent2016, Podgorny2021}, numerous attempts of this kind led to important models for the X-ray astronomical community. For example, the models {\tt PEXRAV} and {\tt PEXRIV} \citep{Magdziarz1995}, {\tt REFLIONX} \citep{Ross1993, Ross1999, Ross2005}, and more recently the {\tt XILLVER} tables \citep{Garcia2010, Garcia2011, Garcia2013} computed with the code {\tt XSTAR} \citep{Kallman2001}, include a detailed computation of ionization structure of the disc. In addition to these, we refer to our previous paper \cite{Podgorny2021}, where completely new local reflection tables suitable for AGNs were presented, assuming a constant density slab, and where we compared them to other attempts of spectral modeling in the literature. These local reflection tables were obtained using 1) the radiative transfer code {\tt TITAN} \cite{Dumont2003} to obtain the ionization structure of a disc and 2) the 3D Monte Carlo code {\tt STOKES} \citep{Goosmann2007,Marin2012,Marin2015,Marin2018} that incorporates the physics of absorption, re-emission and multiple scattering to produce a complete spectropolarimetric output. Therefore, in addition to the spectral tables, our previous study provides an up-to-date unique numerical simulation of the locally reflected polarization quantities. A single-scattering Chandrasekhar's analytical approximation was often used to estimate the reflected polarization degree and angle in the past \citep{Chandrasekhar1960, Dovciak2004, Schnittman2009, Dovciak2011}, which was already proven to provide misleadingly higher polarization degree estimates in the local co-moving frame than the precise numerical simulations \citep{Podgorny2021}.

Using the new local computations from \cite{Podgorny2021}, the aim of this article is to introduce realistic models of the Stokes parameters in X-rays of the AGN inner-accreting region for a distant observer in the lamp-post reflection scenario, which will be already suitable for fitting of real sources with the tool {\tt XSPEC} \citep{Arnaud1996}. We focus on the AGNs rather than XRBs, because in the energy ranges important for the forthcoming polarimetric missions ($\gtrapprox 2$ keV), it is so far easier to simulate the reflection in the local co-moving frame for them due to the expected lower densities ($10^{13}$--$10^{18} \ \textrm{cm}^{-3}$ for AGN discs, $10^{19}$--$10^{25} \ \textrm{cm}^{-3}$ for XRB discs) and lower temperatures ($10^{3.5}$--$10^{6} \ \textrm{K}$ for AGN discs, $10^{5}$--$10^{7.5} \ \textrm{K}$ for XRB discs) with the absence of the direct thermal black-body component in X-rays \citep{Shakura1973, Novikov1973, Reynolds2003, Abramowicz2013, Compere2017, Kubota2018}. Although it is more likely that XRB sources in our Galaxy will in this decade bring more valuable information than the fainter AGNs in the scope of photon-demanding polarimetry \citep{Fabiani2014}, it is more practical to first build a complete simulation kit for AGNs, which can be then easily modified once precise local reflection polarization tables for XRBs are computed, which we plan to do in the future also with the code {\tt STOKES} [the polarization induced by absorption of the thermal radiation in the upper layer of the disc for XRBs has been already recently greatly estimated in \cite{Taverna2020b}, also using the codes {\tt CLOUDY} \citep{Ferland2013, Ferland2017} and {\tt STOKES}, and it was already studied before by e.g. \cite{Dovciak2008}].

Similarly to the {\tt XILLVER} local spectral reflection tables, which have their own relativistic extension {\tt RELXILL} developed \citep{Garcia2014, Dauser2014}, in this paper we introduce the {\tt STOKES} spectral \textit{and} polarization tables tied with the well-tested relativistic {\tt KY} package \citep{Dovciak2004b,Dovciak2011}, covering all of their parametric dependencies, and thus creating a new {\tt XSPEC} compatible model named {\tt KYNSTOKES}. The spectropolarimetric reflection scenario utilizing the {\tt KY} codes was already addressed in \cite{Dovciak2011} in the lamp-post coronal assumption, using the {\tt NOAR} spectral reflection tables for neutral disc and using the Chandrasekhar's approximation for polarization studies. They discussed the role of BH spin, observer's inclination and lamp-post height above the BH with respect to the emergent X-ray polarization. The analytical formulae used for reflection from the disc therein \citep{Chandrasekhar1960} involve only the elastic Rayleigh single scattering, which is greatly improved by this work, presenting Compton multiple-scattering treatment and addition of spectral lines. Compared to the results presented in \cite{Dovciak2011}, and apart from the newly addressed ionization structure of the disc and the numerically simulated local reflection, we also corrected a sign error in \cite{Dovciak2011} concerning the local polarization angle, added arbitrary polarization of the primary source and implemented calculations of the relativistic rotation of the primary polarization angle from the lamp towards a distant observer and from the lamp towards the disc, where we can now interpolate incident polarization in the local frame owing to the new {\tt STOKES} tables. As this paper endeavors to present new theoretical models and to discuss their inherent spectral and polarization features, we plan to address any direct and more sophisticated observational perspectives related to particular targets or instruments in our future studies.

The structure of the text is as follows: Section \ref{model} provides an overview of the physical model assumed, the numerical techniques performed [stressing the novelties with respect to the previous attempt by \cite{Dovciak2011}]. Section \ref{results} is devoted to the obtained spectral and polarization properties for a distant observer. In Section \ref{discussion} we present a comparison with a similar approach previously appearing in the literature, discuss the strengths and weaknesses of the new model, and lay the ground for more extensive, future research. In Section \ref{conclusion} we conclude our analysis.

\section{The model and numerical implementation}\label{model}

\subsection{Physical model of the BH and its accreting structure}

The computation of the results presented in Section \ref{results} is split in three separate stages. First, we take an optically thick slab of constant density in a local frame co-moving with the disc, i.e. without any relativistic effects on the radiation. We use the radiative transfer code {\tt TITAN} to solve for the ionization structure of the illuminated slab iteratively, assuming non-LTE conditions and 500 vertical layers of the disc's plane-parallel atmosphere. In a second step we remain in the local frame and use the same physical setup but in a Monte Carlo simulator {\tt STOKES}. This approach provides a detailed treatment of the polarization, which is not implemented in {\tt TITAN}. Conversely, a Monte Carlo simulation is unable to calculate the disc structure self-consistently. Therefore, in order to take into account for all line and continuum process in {\tt STOKES} inside the disc atmosphere, we implement as input to {\tt STOKES} the information on the temperature, density (constant in our case) and ionization structure (fractional abundance of each element in different states of ionization) per each layer from {\tt TITAN}, which computed the same problem in step 1. After that we average the spatial resolution of {\tt TITAN} to a maximum of 50 vertical layers for computational efficiency of the Monte Carlo method. Combining consistently (i.e. for the same physical problem) a radiative transfer calculus treatment with a Monte Carlo simulator, we arrive at local spectropolarimetric reflection tables \citep[in detail described in][]{Podgorny2021}. These take into account the photon energy redistribution via disc re-processing, being dependent on the local incident and emission angles, the ionization parameter (representing the local bolometric flux and the slab density) and the primary power-law index.

The tables are then used in the third stage, which is the main focus of this paper. We interpolate and integrate them across the accretion disc with the new {\tt KYNSTOKES} code. We assume on-axis spherical corona geometry approximated by a point source as the only primary source of X-ray radiation and we take into account all GR effects (apart from disc's self-irradiation) on radiation that travels between the corona and the disc, between the disc and the observer and between the corona and the observer. Further we consider an equatorial geometrically thin and optically thick Keplerian accretion disc of fixed radial size. Using the {\tt KYNSTOKES} integrator, we then obtain a spatially unresolved total spectropolarimetric output for a distant observer. Multiple scatterings inside the disc are taken into account, but if a photon escapes the geometrically thin disc atmosphere, we do not account for higher order re-processing inside the disc, if the photon's trajectory would be bent such that it would be reflected again off the disc. Photons from the lamp and from the first re-processing in the disc that do not arrive to the distant observer are discarded. Let us now describe each step in more detail.

\subsubsection{Local reflection}\label{local}

In \cite{Podgorny2021} we presented the local X-ray reflection computations that will be primarily used for the results presented in this paper. We assumed the disc to be locally a plane-parallel slab with an electron scattering-dominated atmosphere \citep{Shakura1973, Novikov1973}. We studied semi-infinite and optically thick medium with constant density $n_\mathrm{H} = 10^{15} \textrm{ cm}^{-3}$. The illumination by a primary power-law radiation then creates vertically stratified ionization structure characterized by the ionization parameter in $[\textrm{erg} \cdot \textrm{cm} \cdot \textrm{s}^{-1}]$ \citep[see e.g.][]{tarter1969}
\begin{equation}\label{xi}
	\xi = \dfrac{4 \pi \int F_\mathrm{E} (r) \mathrm{d}E}{n_\mathrm{H}} \textrm{ ,}
\end{equation}
where $F_\mathrm{E}(r) \sim E^{-\Gamma+1}$ is the radiation flux locally received on the surface of the slab. We adopted the typical solar abundance from \cite{Asplund2005} with $A_\mathrm{Fe} = 1.0$, neglecting the presence of dust. Thermal emission was yet excluded from our study, as well as the impact of inverse Compton scattering in the disc re-processing. We neglect any possible primary radiation impact from the opposite side of the disc. Linear polarization of the incident X-rays was allowed, because it is expected at the hot coronal electrons due to Compton up-scattering of thermal photons \citep[see e.g.][]{Haardt1993, Haardt1993b, Schnittman2010}. The emergent spectropolarimetric state was studied after computations of various line and continuum processes inside the disc, most importantly the multiple Compton down-scattering. We refer to \cite{Podgorny2021} for detailed implementation of these assumptions using the codes {\tt TITAN} and {\tt STOKES}.

Three distinct polarization states of the primary radiation were considered: a) unpolarized light with the linear polarization degree $p_0 = 0\%$, b) horizontally polarized light with $p_0 = 100\%$ and the incident polarization angle\footnote{We define the local polarization angle throughout this paper as increasing in the counter-clockwise direction in the incoming view of photon direction with $\chi = 0$ corresponding to a polarization vector oriented along the projected disc's normal to the polarization plane.} $\chi_0 = 90^{\circ}$, c) diagonally polarized light with $p_0 = 100\%$ and $\chi_0 = 45^{\circ}$. The illumination and reflection from the slab was then studied for all possible incident polar angles\footnote{Then $\mu_\mathrm{i}$ represents a cosine of an angle $\delta_\mathrm{i}$ measured from the disc's normal.}, $\delta_\mathrm{i}$, emergent polar angles\footnote{Then $\mu_\mathrm{e}$ represents a cosine of an angle $\delta_\mathrm{e}$ measured from the disc's normal.}, $\delta_\mathrm{e}$, and azimuthal angles\footnote{A relative azimuthal angle measured counter-clockwise at the disc plane between the emission and the incident ray directions projected to the disc's surface.}, $\Phi_\mathrm{e}$. See figure 1 in \cite{Dovciak2011} for a scheme.

Given these assumptions and configurations, the locally reflected flux $F_\mathrm{E}$ in $[\textrm{keV} \cdot \textrm{cm}^{-2} \cdot \textrm{s}^{-1} \cdot \textrm{keV}^{-1}]$ (see \cite{Podgorny2021} for the details on the normalization used) and its polarization degree $p$ and polarization angle $\chi$ can then be studied with respect to energy, the parameters $\xi$, $\Gamma$, $\mu_\mathrm{i}$, $\mu_\mathrm{e}$, $\Phi_\mathrm{e}$, and the variable incident polarization states. We refer to table 1 in \cite{Podgorny2021}, which summarizes the resulting FITS local reflection tables and the complete adopted parametric grid. More than $3.5 \times 10^8$ photons were emitted per each configuration. We proved that the Monte Carlo approach leads into spectral features that match well with other predictions in literature using the equations of radiative transfer. The spectral discrepancies vanish especially near the so-called Compton hump at around 20 keV. Gaining confidence in the spectral results, we discussed the obtained local polarization, which has been for the first time properly numerically simulated for disc re-processing. We demonstrated that in the energy band $2$--$12\,$keV important for the upcoming polarimetric missions, the locally reflected X-rays even for unpolarized primary radiation can reach $p$ up to about $90\%$ in the most favourable parametric configurations. However, with the implementation of multiple scattering, the predicted degree of polarization is generally lower than using the common single-scattering Chandrasekhar's analytical approximation \citep{Chandrasekhar1960}. We obtained constant continuum polarization angle in energy with some variations in spectral lines. The spectral lines being present mostly below $2\,$keV generally tend to depolarize the emission at soft X-rays. We demonstrated that variable incident polarization impacts the local spectral and polarization output heavily, depending on the dominant scattering angle in the medium; nonetheless, if we evenly integrate over the incident and emergent angular space, the impact of variable incident polarization on the spectral part of the emergent radiation is negligible.
\begin{figure*}
	\includegraphics[width=2.\columnwidth]{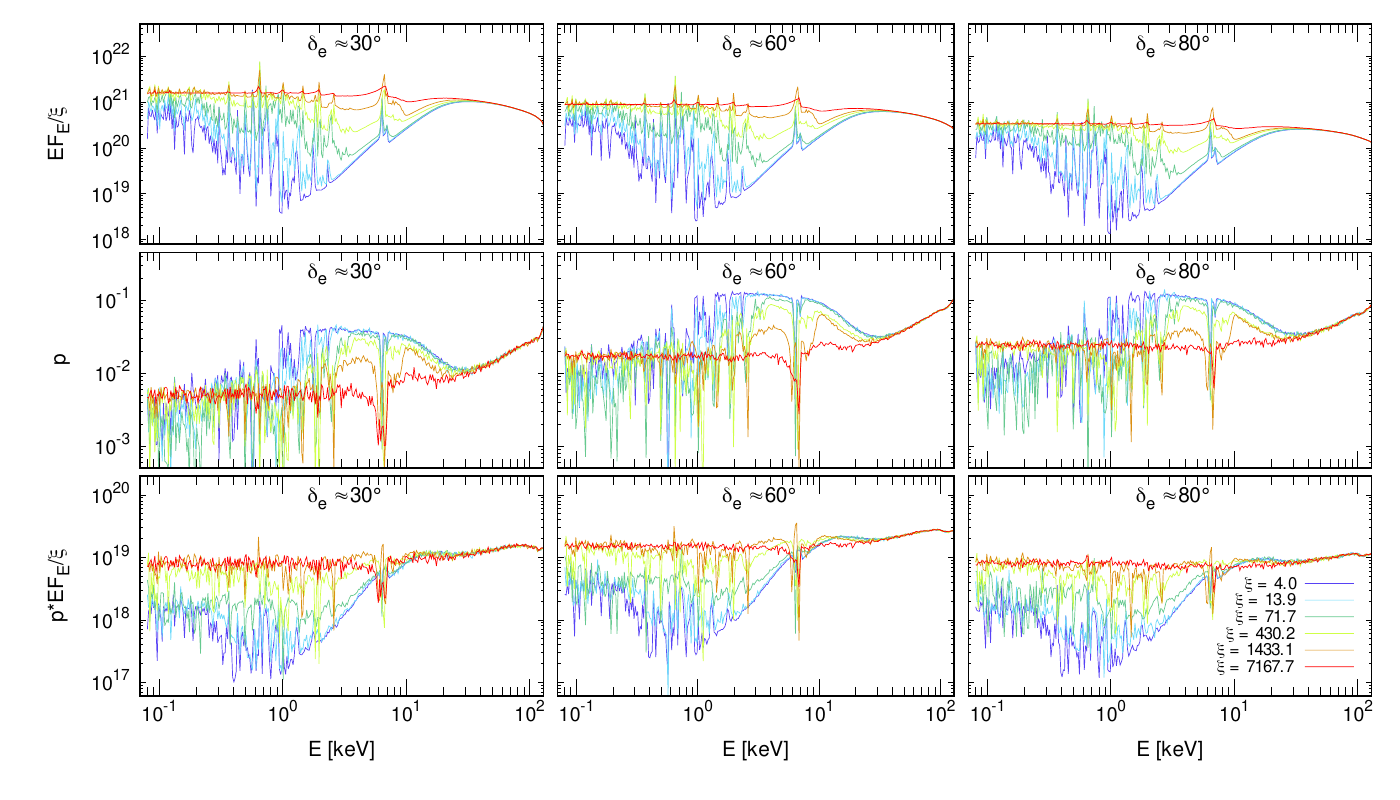}
	\caption{Local reflection spectral results $EF_\mathrm{E}/\xi$ (corrected for primary power-law slope and divided by ionization parameter to unify the amplitude, top panel), polarization degree versus energy (middle panel), and polarized flux (bottom panel) obtained by {\tt STOKES}, integrated in incident inclination angles and emergent azimuthal angles for spectral index $\Gamma = 2.0$. The color code corresponds to various ionization parameters $\xi$. We show the values of $\mu_\mathrm{e} = 0.875, 0.475, 0.175$ (left to right panels, $\delta_\mathrm{e} = 28.96^\circ, 61.65^\circ, 79.92^\circ$) correspondent to the global disc inclinations displayed in the next sections.}
	\label{fig:local_not_in_mue.}
\end{figure*}
\\
\\
Here we would like to provide a couple of new results obtained from our local computations using simple integration in the incident and emergent local angular space for unpolarized primary radiation, in addition to those presented in \cite{Podgorny2021}. The integrated local tables over the angular parameters represent an important middle step towards the analysis of the global results that will be discussed in Section \ref{results}, because they mimic the general relativistic integration over the disc in an even way, i.e. not preferring any direction due to light aberration and light bending effects. Integrating the local tables with an even weight over all cosines of incident angles $\mu_\mathrm{i}$ and emergent azimuthal angles $\Phi_\mathrm{e}$, but not over the emergent inclination angles $\mu_\mathrm{e}$, gives Newtonian insights and represents the dependence on disc inclination for a distant observer. Such examination is also additionally valuable, because it allows better photon statistics\footnote{In \cite{Podgorny2021} we rather summed over the 10 neighboring energy bins for basic discussion of the polarimetric quantities.} in the full available energy resolution of 300 logarithmically spaced bins between 0.1 and 100 keV.

Figure \ref{fig:local_not_in_mue.} provides this integrated version of the local photon-flux in the top panels for $\Gamma = 2$ corrected for the primary power-law slope, i.e. multiplied by $E^2$, and divided by the ionization parameter $\xi$, which sets the amplitude of the spectra. In this way we can inspect the effects of continuum absorption and spectral line profiles with respect to the varying ionization parameter, which is represented by the color code. The selected three values of $\mu_\mathrm{e} = 0.875, 0.475, 0.175$ (from left to right) are the closest available values from the FITS tables that correspond to the global inclination angles discussed in the next sections and in \cite{Dovciak2011}, for which this paper represents a major update. We also show the local polarization degree versus energy for the same set of parameters.

We observe that the integrated polarization degree maintains its behavior with energy and the ionization parameter, as it was already shown and explained in \cite{Podgorny2021} for one generic angular configuration. The enhancement by absorption is still present. Due to the good photon statistics in the highest available resolution, we may notice that the emission lines clearly depolarize. Spectral lines due to photo-absorption or recombination are assumed to be unpolarized. Nevertheless, line polarization may be induced by (multiple) resonant line scattering. Therefore, the corresponding routines in {\tt STOKES} include the treatment worked out and presented in the series of papers by \citet{Lee1994a, Lee1994b, Lee1997}. The competing Auger effect is also implemented. The polarization of the scattered line photon depends on the quantum numbers characterizing the electronic transition of the resonant line. It covers a wide range in strength, but may become as relevant as for Thomson scattering. For example, \cite{Dorodnitsyn2010, Dorodnitsyn2011} showed a notable polarization contribution due to resonance lines below 2 keV in a detailed study of warm absorbers in AGNs. In our study, we obtained a dominance of fluoerescence lines, resulting in depolarization with respect to the Compton scattered continuum. However, if higher spatial and energy resolution were implemented, it could result into some highly polarized resonance lines niether washed out by subsequent multiple scattering before escaping the atmosphere, niether by the relativistic integration of total disc emission. It is also worth mentioning that any current or forthcoming X-ray polarimeter is incapable of resolving individual spectral lines in polarization, especially in case of faint AGN \citep{Fabiani2014}.

The angular superposition sets the average emergent polarization degree to have a maximum of only about $5\%$ to $10\%$ in the total range, although the maxima can be even lower depending on the local ionization of the medium and a particular energy band. We do not show the integrated local polarization angle as it only obtains $\chi \approx 90^{\circ}$ value at all energies ($\lesssim 2^{\circ}$ variation in the total range), apart within unpolarized lines where the polarization position angle is, by definition, undefined. In Figure \ref{fig:local_not_in_mue.} we also show the local polarized flux (i.e. the first two quantities multiplied) for the same set of parameters. It not only supports the antagonistic behavior of the flux and polarization degree with energy, but represents the original Stokes parameters $Q = Ip\cos{2\chi}$ and $U = Ip\sin{2\chi}$ without any sign oscillations (because the continuum of $\chi$ remains constant in energy), and again abstracted away from the slope and amplitude changes.

One can see that the dependency on $\xi$ does not change character with variable $\mu_\mathrm{e}$. The emergent inclination angle, roughly representing the global disc inclination in the Newtonian limit, is expected to decrease the total flux (from left to right), because more absorption takes place for emergent directions closer to the disc plane. If we suppress the aberration of light and general relativistic light bending effects in this sense, the local computations predict higher polarization degrees for higher disc inclination, while the difference is not large above $\delta_\mathrm{e} \approx 60^{\circ}$. The average polarization fraction is about 3 times larger for $\delta_\mathrm{e} \approx 60^{\circ}$ than for $\delta_\mathrm{e} \approx 30^{\circ}$ in the studied energy bands and for the studied $\xi$ values. This generally results into the highest polarized flux for medium emission inclination angles (around $\delta_\mathrm{e} \approx 60^{\circ}$), which suggests the most favourable global inclination for polarization measurements.

\subsubsection{Global model and the relativistic effects}\label{global}

We keep the convention of polarization angle $\chi = 0$ being parallel to the disc axis of symmetry and increasing counter-clockwise when viewing in the photon's incoming direction. For a sketch of the lamp-post model of the accreting structure that we adopt, see figure 1 in \cite{Dovciak2011}. Similarly to this work, we assume an elevated isotropically emitting point-like source at height $h$ on the rotational axis above the equatorial plane, illuminating the geometrically thin disc in this plane with a power-law radiation with some observed primary isotropic flux $L_{\textrm{X}}/L_{\textrm{Edd}}$ at $2-10 \textrm{ keV}$ in the units of Eddington luminosity $L_{\textrm{Edd}}$. The disc is assumed to have a Keplerian velocity profile and a constant radial density profile (in this paper $n_\textrm{H} = 10^{15} \ \textrm{cm}^{-3}$). The system can be inclined at general inclination $i$ and with arbitrary orientation with respect to the distant observer. We assume the emission region to be unobscured, i.e. we do not aim to model any other outer components of the accreting structure, such as jets, the broad-line regions (BLRs), narrow-line regions (NLRs), polar winds, or a dusty torus.

We prescribe the inner and outer disc radius $r_{\mathrm{in}}$ and $r_{\mathrm{out}}$ (in the results presented $r_{\mathrm{in}}$ is fixed at the inner-most circular orbit (ISCO), $r_{\mathrm{out}} = 400 \ GM/c^2$) and compute the geodesics from the lamp to the disc in Kerr space-time with arbitrary normalized black-hole spin $a$ and black-hole mass $M_{\mathrm{BH}}$. The disc is assumed to be geometrically thin in the equatorial plane but optically thick, i.e. we consider only photons coming from the source or disc directly to the observer and no source on the opposite side and we neglect the disc self-irradiation \citep[see e.g.][]{Schnittman2010, Dauser2022}. The effects of radiation returning to the disc after being re-processed once could affect the resulting polarization picture though, which is discussed in Section \ref{future_improvements} as one of the critical simplifications of this work. Otherwise all special and general relativistic effects are included, such as the aberration of light, Doppler and gravitational redshifts, light bending and rotation of the polarization plane. We then superpose geodesics from the disc to the distant observer (\textit{reflected-only radiation}) and from the lamp to the distant observer (\textit{primary radiation}) to produce a combined spectro-polarimetric outcome at infinity (\textit{total radiation}).

\subsection{Numerical models}\label{models}

As a major improvement with respect to \cite{Dovciak2011}, we incorporated the new local reflection FITS tables from \citep{Podgorny2021} that were computed with the codes {\tt TITAN} and {\tt STOKES}. These include numerically simulated polarization properties of the disc reflection as well as detailed calculations of the vertical ionization disc structure. Thus polarization properties of the reflection from ionised discs can be studied. Further, the reflection tables were computed for three independent incident polarization states defined by the polarization degree and angle, ($p$, $\chi$) = (0, --), \text{(1, $\pi/2$)}, (1, $\pi/4$). The disc response computed by the {\tt STOKES} code is a linear transformation of the incident Stokes parameters, $I$, $Q$ and $U$, and therefore the disc response to an arbitrary incident polarization state can be computed from this basis. Since GR effects do not change the polarization degree between the primary source and the disc and they change the polarization angle by $\chi_{\rm d}$, one can finally express the reflected Stokes parameters (commonly denoted here as $S$) at each point as 
\begin{equation}\label{Sincident}
\begin{split}
    S(p_0, \chi_0+\chi_{\rm d}) = & \, S(0,-) + p_0\{ [S(0,-)-S(1,\pi/2)]\cos{2(\chi_0+\chi_{\rm d})} \\
    &\quad + [S(1,\pi/4) - S(0,-)]\sin{2(\chi_0+\chi_{\rm d})} \}.
\end{split}
\end{equation}
where the primary polarization state is defined by the polarization degree, $p_0$, and the polarization angle, $\chi_0$. Naturally, when we compute the observed polarization, full general-relativistic computations of the change of the polarization angle from the lamp towards a distant observer and towards the disc as well as from the disc towards the observer are implemented, see Appendix \ref{calculations} for more details. Lastly, note that with respect to \cite{Dovciak2011}, we have also corrected a sign error in the definition of the local polarization angle of the disc emission, see Appendix \ref{loc_error} for more details. Because we plan to rework the entire discussion of polarization for a distant observer in Section \ref{results}, using the new local reflection tables from \cite{Podgorny2021} and including other improvements in the KY codes since \cite{Dovciak2011}, this correction does not propagate to the main results of this paper. We name the new model as {\tt KYNSTOKES}, with reference to the Monte Carlo code, {\tt STOKES}, that was used to compute the local reflection polarization tables. See the data availability section for a link to external repository of {\tt KYNSTOKES}, including user instructions.

\section{Results for a distant observer}\label{results}

\begin{table*}
    \centering
	\caption{Summary of all figures provided in Sections \ref{results} and \ref{discussion} that concern discussion of the {\tt KYNSTOKES} results and comparisons for the global model viewed from infinity. We stress in bold the most visible distinction of the curves in each figure, which is given by the color code. All figures and curves within are for the case $\Gamma = 2$, if not mentioned otherwise. We distinguish the disc ionization by the $M_\textrm{BH}$ and $L_\textrm{X}/L_\textrm{Edd}$ parameters in the global model (see the text further).}
	\resizebox{\textwidth}{!}{%
	\begin{threeparttable}
    \begin{tabular}{lllllll}
    \hline \hline
    $y$-axis $\rightarrow$                                                                                                                                 & \multicolumn{2}{c}{Intensity (normalized to value at 50 keV)}                                             & \multicolumn{2}{c}{Polarization degree}                                         & \multicolumn{2}{c}{Polarization angle}                                            \\ 
    $x$-axis (and curves displayed) $\downarrow$                                                                                                                                 & \textbf{Reflected-only} $EF_\mathrm{r,E}$                     & \textbf{Total} $EF_\mathrm{E}$                               & \textbf{Reflected-only} $p_\mathrm{r}$              & \textbf{Total} $p$                       & \textbf{Reflected-only} $\chi_\mathrm{r}$              & \textbf{Total} $\chi$                             \\ \hline \hline
    vs. $E$ (for \textbf{3 disc ionizations}, $2\times h$,                             & \ref{fig:K3_R_unpol_I.} & \ref{fig:K3_RP_unpol_I.} &     \ref{fig:K3_R_unpol_pdeg.}                          &                  \ref{fig:K3_RP_unpol_pdeg.}             &           \ref{fig:K3_R_unpol_pang.}                    &       ($=$ \ref{fig:K3_R_unpol_pang.})                       \\  \ \ \ \ $2\times a$, $3\times i$, unpolarized primary)                                   &  &  &                           &                              &                         &       
    
    \\ \hline
    vs. $h$ and $i$ (for neutral disc, $2\times a$,                                          &                                     &                                     &                               &             \ref{fig:fig9_tot_neutral.}                  &                               & \ref{fig:fig10_tot_neutral.}\tnote{b} \\  \ \ \ \ \textbf{5 energy bands}, unpolarized primary)                                           &                                     &                                     &                               &                       &                               &  \\ \hline
    vs. $h$ and $i$ (for highly ionized disc, $2\times a$,                                    &                                     &                                     &                               & \ref{fig:fig9_tot_ionized.}\tnote{a} &                               &        \ref{fig:fig10_tot_ionized.}                  \\    \ \ \ \  \textbf{5 energy bands}, unpolarized primary)                                    &                                     &                                     &                               &  &                               &          \\ \hline
    vs. $E$ (for neutral disc, $2 \times h$, one choice of $(a,i)$                                &                                     &                                     &                   \ref{fig:K3A_R_polar_n_pdeg_unique.}          &                  \ref{fig:K3A_RP_polar_n_pdeg_unique.}            &      \ref{fig:K3A_R_polar_n_pang_unique.}                                &                                    \\  \ \ \ \ \textbf{3 incident polarization states})                                   &                                     &                                     &                           &                             &                             &     \\ \hline
    vs. $E$ (for neutral disc, $2\times h$, $2\times a$, $3\times i$,                             &                                     &                                     &  &  &  &       \ref{fig:K3A_RP_polar_n_pang_mainbody.}                             \\ \ \ \ \   \textbf{3 incident polarization states})                               &                                     &                                     &  &  &  &           \\ \hline
    vs. $E$ (for \textbf{2 different reflection tables}, neutral disc,  &                                     &                                     & \ref{fig:K5_p_Ln.}\tnote{c}                & \ref{fig:K5_p_Ln_RP.}\tnote{c}                & \ref{fig:K5_Psi_Ln.}\tnote{c}                & \ref{fig:K5_Psi_Ln_RP.}\tnote{c}          \\     \ \ \ \  $1\times h$, $2\times a$, $3\times i$, 3 incident polarization states) &                                     &                                     &             &             &          &       \\ \hline \hline
    \end{tabular}
    \begin{tablenotes}
    \item [a] Is also re-created for $\Gamma = 3$ in Figure \ref{fig:fig9_tot_ionized_30.}.
    \item [b] Is also re-created for $p_0 = 0.01, \chi_0 = 90^\circ$ in Figure \ref{fig:fig10_tot_neutral_P2.}.
    \item [c] The $x$-axis range is shrinked to $E\in [1,100]\,$keV.
    \end{tablenotes}
    \end{threeparttable}
    }
    \label{summary_figures}
\end{table*}
In order to discuss the spectral and polarization properties obtained by {\tt KYNSTOKES} clearly, for each set of parameters, we will first elaborate on the reflected-only emission and then show the total emission including the primary radiation. We will first discuss the spectra, polarization degree and polarization angle versus energy for unpolarized primary radiation and $\Gamma = 2$ and only in the end add a commentary on the effect of non-zero incident polarization and variable photon-index.
We show examples for two cases of the black-hole spin, non-rotating Schwarzschild black hole with $a = 0$ and maximally rotating Kerr black hole with $a = 1$, three inclinations, $i = 30^{\circ}, 60^{\circ}$ and $80^{\circ}$\footnote{Although the high inclinations are not plausible for direct observations due to expected obscuration, the obtained spectro-polarimetric properties might be useful for precise modeling of e.g. torus illumination \citep{Marin2018b, Marin2018c}, therefore we show these examples as well. In addition, by this choice we allow a direct comparison with the similar selection of parameters in \cite{Dovciak2011}.}, for the energy range of $0.1-100\,$keV with 300 energy bins following the resolution of the local reflection tables, although {\tt KYNSTOKES} allows arbitrary binning. Note that while the $y$-range for all shown quantities (relative flux, polarization degree and angle) is fixed within panels of each figure for better comparison, it changes between the figures for better readability. Table \ref{summary_figures} summarizes the order and content of all figures in the main paper body.

\subsection{Intensity}

\begin{figure*}
	\includegraphics[width=2.\columnwidth]{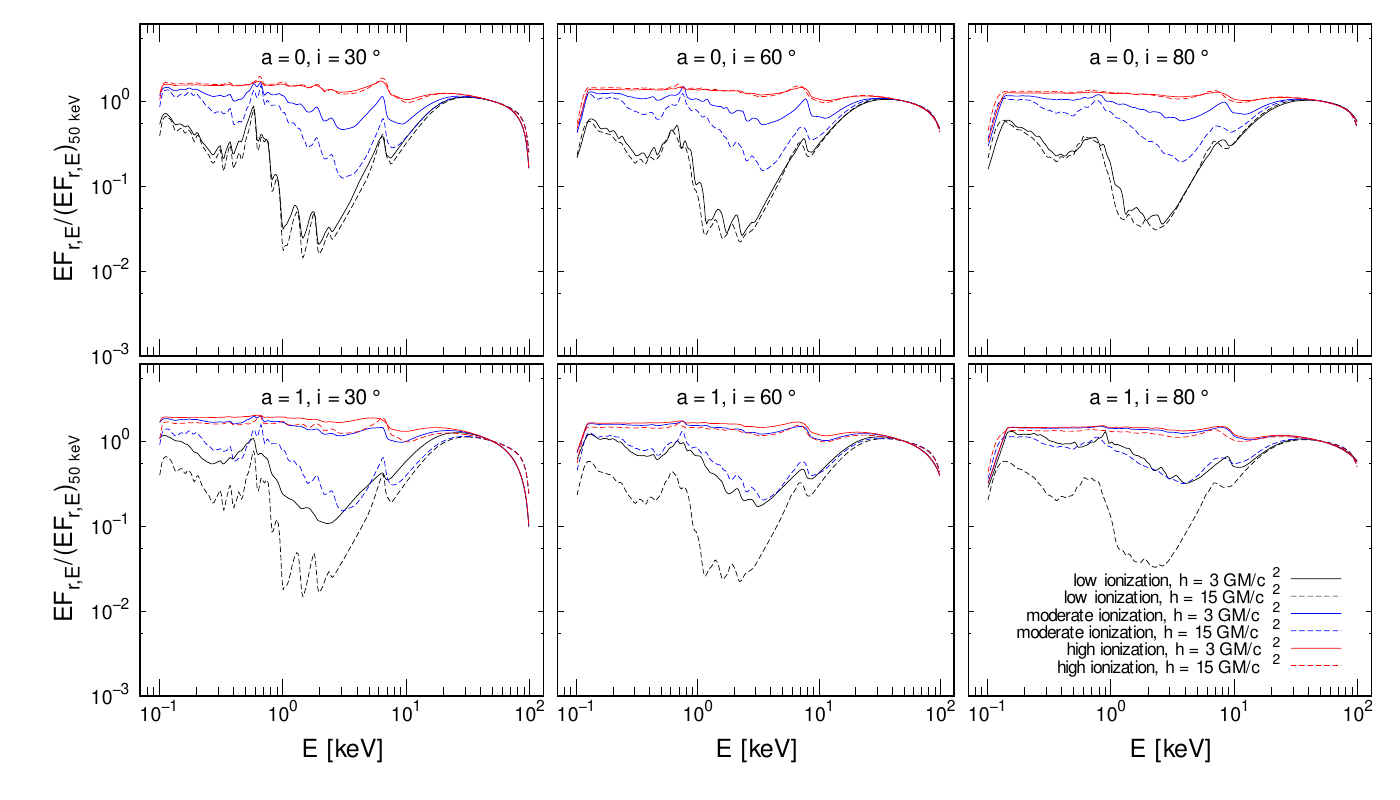}
	\caption{The reflected-only spectra of the accretion disc for distant observer, $EF_\mathrm{r,E}$, normalized to value at 50 keV, obtained by {\tt KYNSTOKES} for black-hole spins $a = 0$ (top) and $a = 1$ (bottom), disc inclinations $i = 30^{\circ}$ (left), $i = 60^{\circ}$ (middle) and  $i = 80^{\circ}$ (right), $\Gamma = 2$ and unpolarized primary radiation, using the {\tt STOKES} local reflection model in the lamp-post scheme. We show cases of two different heights of the primary point-source above the disc $h = 3 \textrm{ } GM/c^2$ (solid lines) and $h = 15 \textrm{ } GM/c^2$ (dashed lines), and neutral disc (for $M_{\textrm{BH}} = 1\times 10^8\,M_{\odot}$ and observed 2--10 keV flux $L_{\textrm{X}}/L_{\textrm{Edd}} = 0.001$, black lines), moderately ionized disc (for $M_{\textrm{BH}} = 3\times 10^6\,M_{\odot}$ and observed 2--10 keV flux $L_{\textrm{X}}/L_{\textrm{Edd}} = 0.01$, blue lines) and highly ionized disc (for $M_{\textrm{BH}} = 1\times 10^5\,M_{\odot}$ and observed 2--10 keV flux $L_{\textrm{X}}/L_{\textrm{Edd}} = 0.1$, red lines).}
	\label{fig:K3_R_unpol_I.}
\end{figure*}
\begin{figure*}
	\includegraphics[width=2.\columnwidth]{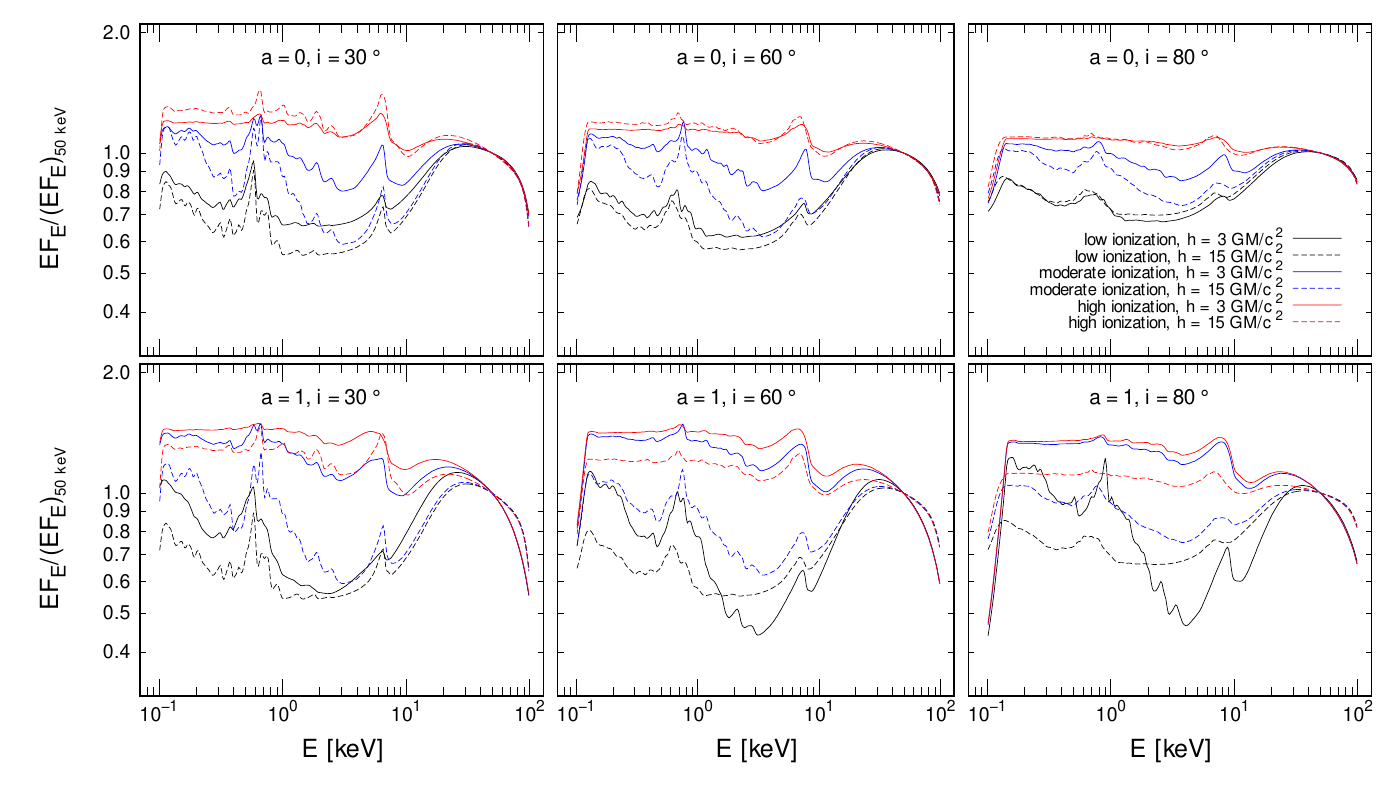}
	\caption{The total spectra of the accretion disc for distant observer, $EF_\mathrm{E}$, normalized to value at 50 keV, for the same parametric setup as in Figure \ref{fig:K3_R_unpol_I.}, displayed in the same manner.}
	\label{fig:K3_RP_unpol_I.}
\end{figure*}
Figures \ref{fig:K3_R_unpol_I.} and \ref{fig:K3_RP_unpol_I.} represent the reflected-only and total spectra for a selection of different ionizations of the disc, 
corona heights, BH spins and system inclinations. We show the spectra normalized to the value of each curve at $50\,$keV and we multiply by $E^\Gamma$ in order to ignore the amplitude and slope changes and in order to show absorption and spectral features more clearly. The disc integrated reflection spectra show the usual features such as the Fe K$\alpha$ line at $6$--$7 \textrm{ keV}$, a forest of lines below $3\,$keV, and the Compton hump above $10\,$keV, that are relativistically smeared, especially in the configurations where more radiation reaches the inner-most parts of the disc (i.e. for lower primary source heights and higher black-hole spins) and for higher inclinations. 
Note that in case of the total observed emission, these features are further diluted by the continuum power-law, Figure \ref{fig:K3_RP_unpol_I.}.
This general spectral behavior obtained by {\tt KYNSTOKES} using the {\tt STOKES} local tables resembles well the predictions previously appearing in the literature \citep[see e.g.][]{Garcia2014}. It was checked that high lamp-post heights represent Newtonian limits on light bending effects and angular aberration (approaching the results shown in Figure \ref{fig:local_not_in_mue.}).

Typically the blurred Fe K$\alpha$ line in the total spectrum serves as a powerful feature for fitting inclination, spin and disc's ionization in spectra of AGN and XRBs \cite[see e.g.][]{Vaughan2004,Dovciak2004, Miniutti2004, Seward2010} due to the fact that its shape strongly depends on these parameters through the GR effects, e.g. the relativistic broadening of the line is most extreme in cases when more radiation hits innermost regions of the disc close to the black hole.
We refer to \cite{Cunningham1975, Dovciak2011, Dauser2014, Dovciak2014} for an extensive discussion of these effects. Here we just shortly summarize the main features:
\begin{enumerate}
    \item the emission angle is not equal to the observer's inclination and varies with the position on the disc,
    \item the emission from different areas of the disc is shifted in energy and amplified or suppressed by GR effects in a non-axisymmetric way,
    \item the illumination pattern and thus ionisation of the disc changes with primary source height and thus changes the disc response\footnote{In this paper, the energy dependence differs substantially in comparison with \cite{Dovciak2011}, as ionization of the disc is properly treated.},
    \item the relativistic rotation of the polarization angle depends on inclination, black-hole spin, and lamp-post height\footnote{This is also a new effect to be considered, as we have now implemented the possibility of a polarized primary towards the disc and towards the observer.}, 
    \item the BH spin shrinks the inner disc radius and changes light bending effects in the vicinity of the black hole.
\end{enumerate}

The spectrum for a distant observer depends strongly on the ionization of the accretion disc. The ionization in the global model is mainly influenced by the BH mass and
intrinsic luminosity of the primary source. The latter increases the ionization parameter, while the former diminishes ionization because it changes the distance in physical units between the source and the disc.
In a more realistic model, we would have to account also for the effect of BH mass on the disc density (directly involved in the definition of the ionization parameter (\ref{xi})) and that the disc density changes with radial coordinate. In our examples we chose three realistic combinations of $M_{\textrm{BH}}$ and $L_{\textrm{X}}/L_{\textrm{Edd}}$ to mimic (almost) neutral, moderately ionized and highly ionized disc. 
The disc ionization (dependent on radius) creates a specific spectral signature for a distant observer, and contributions of differently ionised parts of the disc to the total observed spectrum will be dependent on the corona height and BH spin.

\subsection{Polarization degree}

\begin{figure*}
	\includegraphics[width=2.\columnwidth]{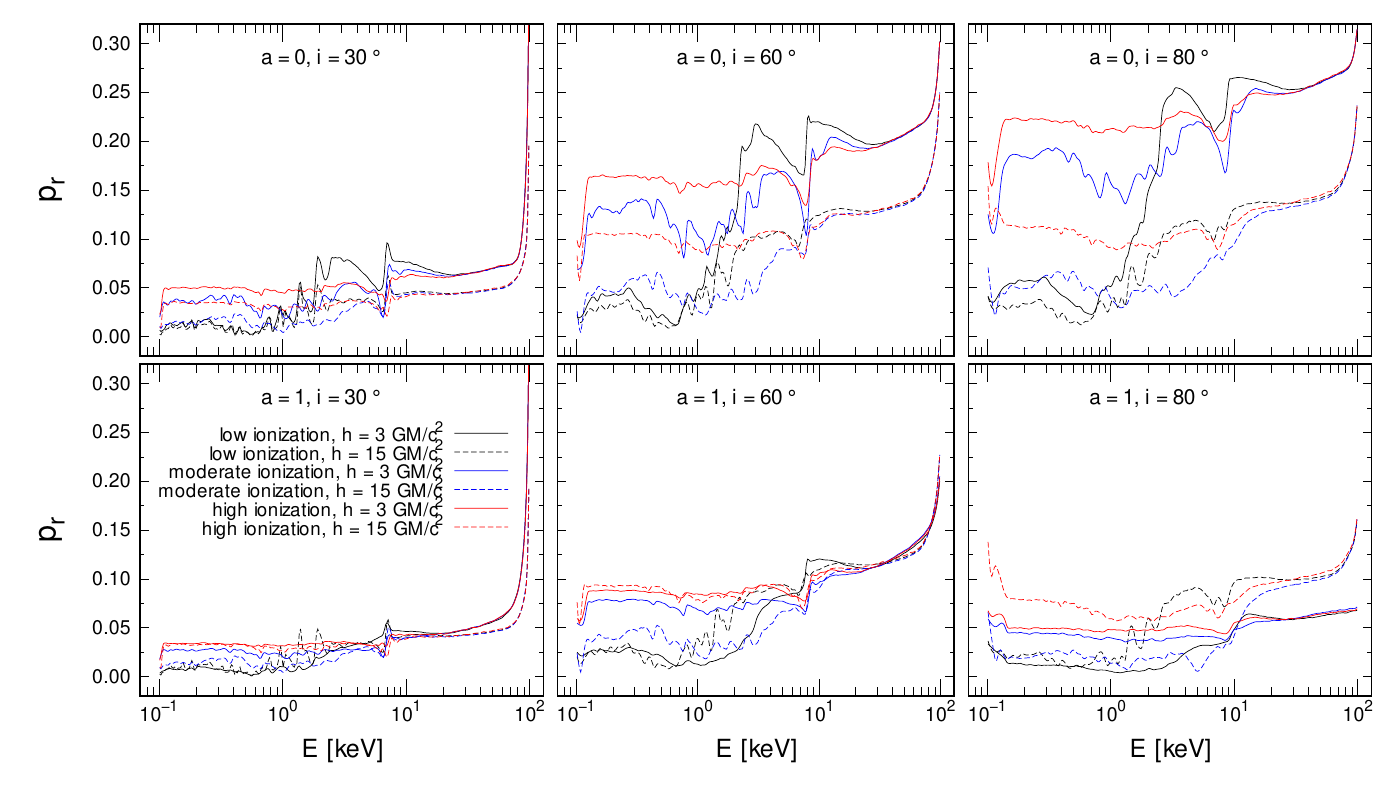}
	\caption{The reflected-only polarization degree, $p_\mathrm{r}$, versus energy for the same parametric setup as in Figure \ref{fig:K3_R_unpol_I.}, displayed in the same manner.}
	\label{fig:K3_R_unpol_pdeg.}
\end{figure*}
\begin{figure*}
	\includegraphics[width=2.\columnwidth]{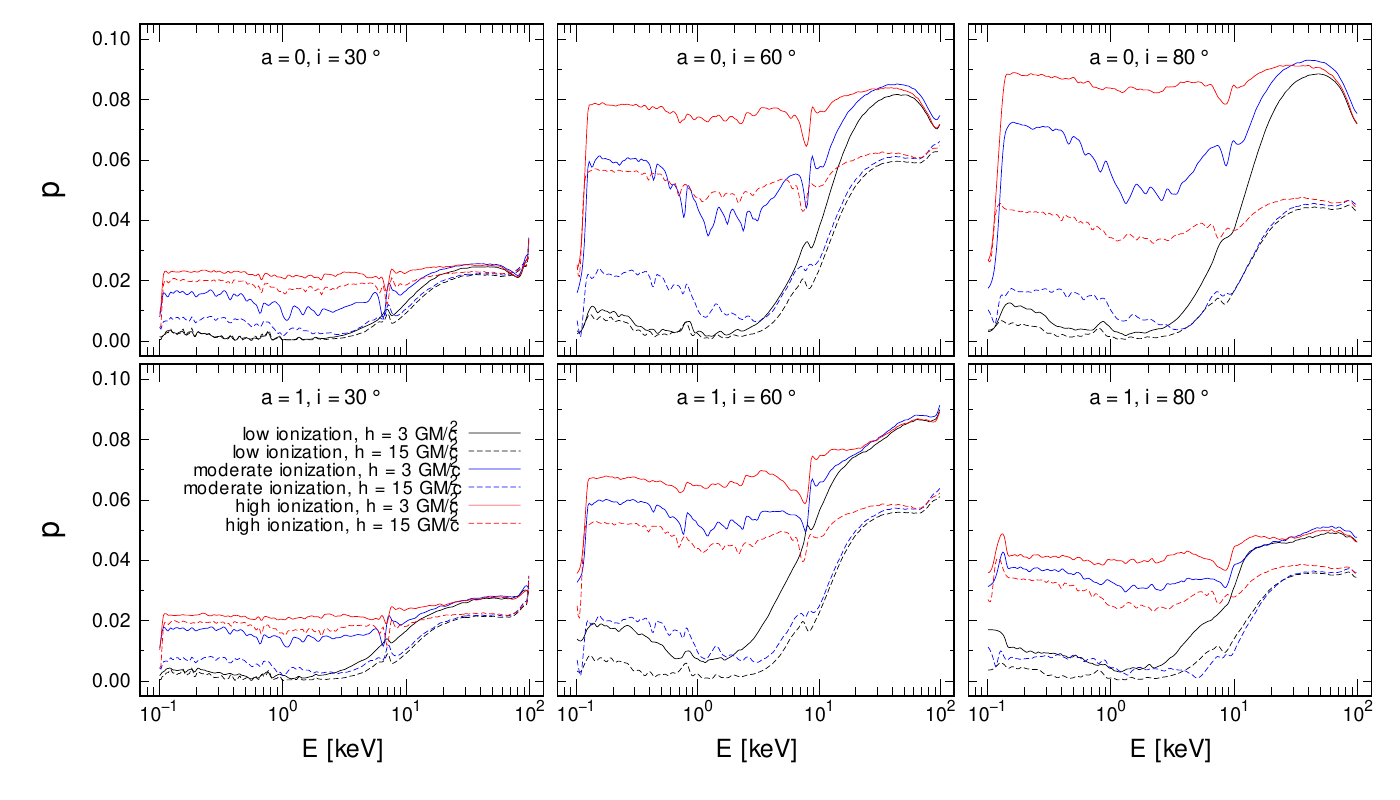}
	\caption{The total polarization degree, $p$, versus energy for the same parametric setup as in Figure \ref{fig:K3_R_unpol_I.}, displayed in the same manner.}
	\label{fig:K3_RP_unpol_pdeg.}
\end{figure*}
Figures \ref{fig:K3_R_unpol_pdeg.} and \ref{fig:K3_RP_unpol_pdeg.} represent the polarization degree with energy for the reflected-only and total emission, respectively. These results are unique, because the local reflection tables from \cite{Podgorny2021} represent the first numerical estimate of locally induced polarization by disc reflection, including proper treatment of multiple Compton scatterings and ionization. We show the same parametric configurations as on the spectral Figures \ref{fig:K3_R_unpol_I.} and \ref{fig:K3_RP_unpol_I.}. One may now estimate to what extent disc reflection can enhance the polarization degree in the AGN inner-accreting region in the lamp-post scheme. Global reflected-only polarization degree can reach up to 25\% in the Compton hump region in the most favourable configurations. The polarization degree induced by reflection at infinity is of course lower than in the local co-moving frame. This was already indicated by the magnitude of the local polarization degree, $p(E)$, integrated over all emission angles shown in Figure \ref{fig:local_not_in_mue.} that is much lower when compared to the results for particular scattering geometry provided in \cite{Podgorny2021}. The total polarization degree can reach up to 9\% in the Compton hump region in the most favourable configurations. This illustrates the depolarization effect of the completely unpolarized primary emission when added to the highly polarized reflected component.

We refer to Table \ref{obs_p} for the predicted average total polarization degree values in the energy ranges of the forthcoming main X-ray polarimetric missions for a few generic parametric selections. The presented model with the {\tt STOKES} local reflection tables has much higher energy resolution than any currently functioning or forthcoming X-ray polarimeter [in 2--8 keV for IXPE \citep{Weisskopf2022} and eXTP \citep{Zhang2016, Zhang2019}, in 15--80 keV for XL-Calibur \citep{Abarr2021}...] could achieve for such faint sources as AGNs, in order to e.g. apply the predicted line behavior for polarimetric fitting of AGN properties. We plan to provide a deep discussion of observational prospects by {\tt KYNSTOKES} in a future paper, including data simulations for some particular X-ray polarimetric missions.

\begin{table}
    \centering
	\caption{The average total polarization degree $100\cdot \overline{p} \, [\%]$ between 2--8 keV (the IXPE or eXTP mission range) and 15--80 keV (the XL-Calibur balloon experiment range) for unpolarized primary radiation and $\Gamma = 2$. We show cases of neutral disc ($M_{\textrm{BH}} = 1\times 10^8\,M_{\odot}$ and observed 2--10 keV flux $L_{\textrm{X}}/L_{\textrm{Edd}} = 0.001$, left cells in black) and highly ionized disc ($M_{\textrm{BH}} = 1\times 10^5\,M_{\odot}$ and observed 2--10 keV flux $L_{\textrm{X}}/L_{\textrm{Edd}} = 0.1$, right cells in red) and various combinations of lamp-post heights $h$, black-hole spins $a$ and disc inclinations $i$.}
	\begin{adjustbox}{max width=\columnwidth}
    \begin{tabular}{cllllllll}
    \hline \hline
                                                     &                          &       & \multicolumn{2}{l}{$i = 30^{\circ}$}                        & \multicolumn{2}{l}{$i = 60^{\circ}$}                        & \multicolumn{2}{l}{$i = 80^{\circ}$}                        \\ \hline
                                                     &                          & $a = 0$ & {\color[HTML]{333333} 0.33} & {\color[HTML]{CB0000} 2.17} & {\color[HTML]{000000} 1.03} & {\color[HTML]{CB0000} 7.47} & {\color[HTML]{000000} 1.03} & {\color[HTML]{CB0000} 8.40} \\ \cline{3-9} 
                                                     & $h = 3 \textrm{ } GM/c^2$  & $a = 1$ & {\color[HTML]{333333} 0.55} & {\color[HTML]{CB0000} 2.15} & {\color[HTML]{000000} 2.12} & {\color[HTML]{CB0000} 6.65} & {\color[HTML]{000000} 1.01} & {\color[HTML]{CB0000} 4.08} \\ \cline{2-9} 
                                                     &                          & $a = 0$ & {\color[HTML]{333333} 0.24} & {\color[HTML]{CB0000} 1.78} & {\color[HTML]{000000} 0.81} & {\color[HTML]{CB0000} 4.97} & {\color[HTML]{000000} 0.43} & {\color[HTML]{CB0000} 3.50} \\ \cline{3-9} 
    \textbf{IXPE and eXTP (2--8 keV)}       & $h = 15 \textrm{ } GM/c^2$ & $a = 1$ & {\color[HTML]{333333} 0.24} & {\color[HTML]{CB0000} 1.73} & {\color[HTML]{000000} 0.59} & {\color[HTML]{CB0000} 4.59} & {\color[HTML]{000000} 0.36} & {\color[HTML]{CB0000} 2.57} \\ \hline \hline
                                                     &                          & $a = 0$ & {\color[HTML]{333333} 2.29} & {\color[HTML]{CB0000} 2.47} & {\color[HTML]{000000} 7.48} & {\color[HTML]{CB0000} 8.24} & {\color[HTML]{000000} 7.95} & {\color[HTML]{CB0000} 9.01} \\ \cline{3-9} 
                                                     & $h = 3 \textrm{ } GM/c^2$  & $a = 1$ & {\color[HTML]{333333} 2.51} & {\color[HTML]{CB0000} 2.62} & {\color[HTML]{000000} 7.78} & {\color[HTML]{CB0000} 7.97} & {\color[HTML]{000000} 4.63} & {\color[HTML]{CB0000} 4.80} \\ \cline{2-9} 
                                                     &                          & $a = 0$ & {\color[HTML]{333333} 2.03} & {\color[HTML]{CB0000} 2.24} & {\color[HTML]{000000} 5.44} & {\color[HTML]{CB0000} 6.10} & {\color[HTML]{000000} 3.99} & {\color[HTML]{CB0000} 4.60} \\ \cline{3-9} 
    \textbf{XL-Calibur (15--80 keV)}       & $h = 15 \textrm{ } GM/c^2$ & $a = 1$ & {\color[HTML]{333333} 1.98} & {\color[HTML]{CB0000} 2.18} & {\color[HTML]{000000} 5.11} & {\color[HTML]{CB0000} 5.70} & {\color[HTML]{000000} 3.25} & {\color[HTML]{CB0000} 3.72} \\ \hline \hline
    \end{tabular}
    \end{adjustbox}
    \label{obs_p}
\end{table}

The general pattern of the reflected polarization degree with energy for a distant observer resembles the pattern expected from the local reflection tables in \cite{Podgorny2021} or their integration in angular space in Figure \ref{fig:local_not_in_mue.}. The role of the ionization parameter $\xi$ is again represented by the combination of $M_{\textrm{BH}}$ and $L_{\textrm{X}}/L_{\textrm{Edd}}$ that mostly affect the disc overall ionization. in Figures \ref{fig:K3_R_unpol_pdeg.} and \ref{fig:K3_RP_unpol_pdeg.} we can clearly see depolarization of lines resulting from the local reflection tables. This is mostly prominently on the Fe line complex, where you can note that the decrease of polarization degree has inverse line profile with the increase of flux in the spectra, including the relativistic smearing.

Due to relativistic effects, both the emission angle and the change of polarization angle vary across the disc. In some cases this can lead to enhancement of the polarization degree, rather than further depolarization with respect to the uniform integration in angular space \citep[for detailed discussion, see e.g.][]{Dovciak2011}. This is seen also in our latest results by comparison of amplitudes of the reflected polarization degree curves in Figures \ref{fig:local_not_in_mue.} and \ref{fig:K3_R_unpol_pdeg.}. For example, the inclination of $60^\circ$ and low BH spin for the highly ionized disc may lead to energy averaged polarization degree around 15\% compared to the maximum ~10\% obtained by the simple Newtonian integration.

\begin{figure*}
	\includegraphics[width=2.\columnwidth]{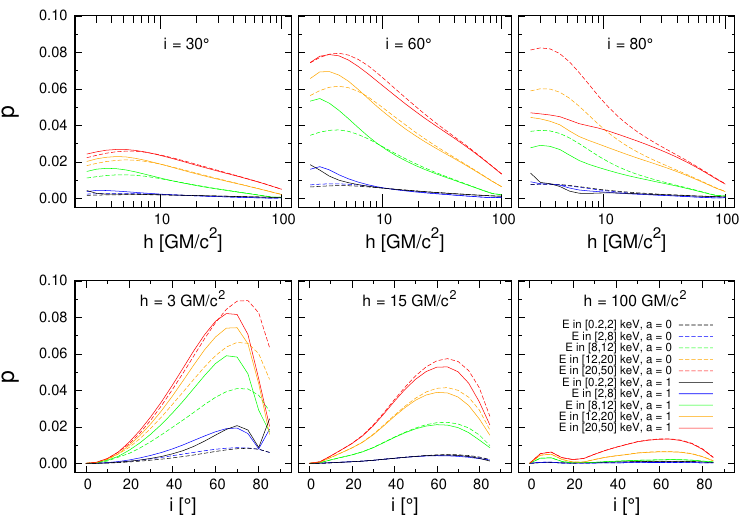}
	\caption{Top panel: the total average polarization degree, $p$, versus height of the primary point-source above the disc for disc inclinations $i = 30^{\circ}$ (left), $i = 60^{\circ}$ (middle) and  $i = 80^{\circ}$ (right), $\Gamma = 2$, unpolarized primary radiation, $M_{\textrm{BH}} = 1\times 10^8\,M_{\odot}$ and observed 2--10 keV flux $L_{\textrm{X}}/L_{\textrm{Edd}} = 0.001$, i.e. neutral disc. We show cases of two black-hole spins $a = 0$ (dashed) and $a = 1$ (solid) and for energy bands $E \in [0.2, 2]$ keV (black),  $E \in [2, 8]$ keV (blue), $E \in [8, 12]$ keV (green), $E \in [12, 20]$ keV (orange), and $E \in [20, 50]$ keV (red). Bottom panel: the total average polarization degree, $p$, versus disc inclination for heights of the primary point-source above the disc $h = 3 \textrm{ } GM/c^2$ (left), $h = 15 \textrm{ } GM/c^2$ (middle) and  $h = 100 \textrm{ } GM/c^2$ (right), for the same configuration as on the top panel.}
	\label{fig:fig9_tot_neutral.}
\end{figure*}
\begin{figure*}
	\includegraphics[width=2.\columnwidth]{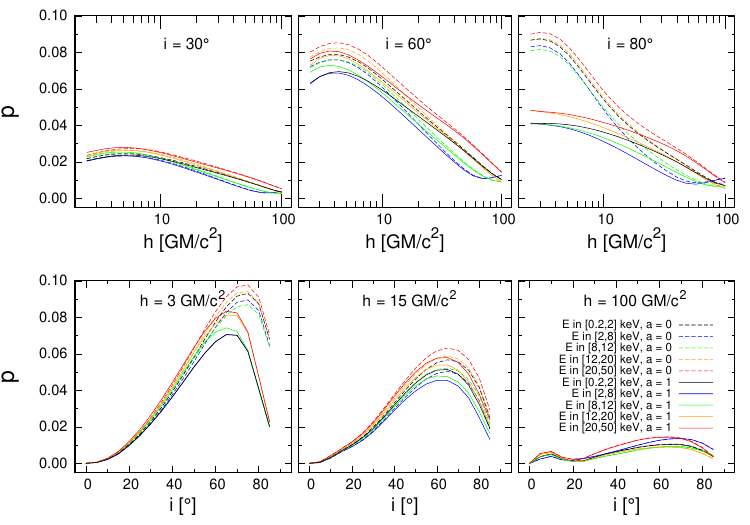}
	\caption{The total average polarization degree, $p$, versus energy for ionized disc ($M_{\textrm{BH}} = 1\times 10^5\,M_{\odot}$ and observed 2--10 keV flux $L_{\textrm{X}}/L_{\textrm{Edd}} = 0.1$), otherwise for the same parametric setup as in Figure \ref{fig:fig9_tot_neutral.}, displayed in the same manner.}
	\label{fig:fig9_tot_ionized.}
\end{figure*}
The dependence of the polarization degree on height and inclination for two BH spin values and different energy bands in the full general-relativistic regime is shown in Figures \ref{fig:fig9_tot_neutral.} and \ref{fig:fig9_tot_ionized.} for a neutral and ionized disc, respectively. These can be directly compared with the results presented in \cite{Dovciak2011}.
Comparing the energy dependence on both of these figures, one can clearly see the energy flattening of $p(E)$ for the ionized case as opposed to the neutral case. For both of these cases, the average polarization degree rises with energy in all configurations, because the harder the detected X-rays are, the more we register purely Compton-scattered photons carrying highly polarized signal.

Figures \ref{fig:fig9_tot_neutral.} and \ref{fig:fig9_tot_ionized.} for the total emission suggest a non-monotonic behavior of $p$ with inclination. It is important to model the polarization also for high inclinations precisely, because the AGN torus polarization models have to account with a possibly polarized illumination \citep{Marin2018b, Marin2018c}. The highest polarized fraction of the X-rays at infinity in our model is generally reached at about $i \approx 65^{\circ}$.
The primary radiation that reaches the observer directly is almost unaffected by the inclination. In Figure \ref{fig:local_not_in_mue.} we already showed that the rise (from $i \approx 30^{\circ}$ to $i \approx 60^{\circ}$) and some flattening (from $i \approx 60^{\circ}$ to $i \approx 80^{\circ}$) of the polarization degree occurs in plain integration of the local tables in the ($\mu_\mathrm{i}$,$\Phi_\mathrm{e}$) space. Thus, we conclude that the behavior of $p$ with $i$ is a combination of averaging over different local geometries and global relativistic effects. For the highest inclinations, the polarized fraction decreases, because the reflection fraction is lower with respect to the unpolarized primary and because more configurations of geometry of scattering are included. The lamp-post with fixed height does not change the incident reflection angles for different disc inclination, while the emission angles vary due to light-bending. The critical point (with photons that reach the observer being emitted in the normal direction to the disc) may be located very close behind the BH from the observer's point of view at high inclinations. Since the reflected flux is the highest from the inner-most disc regions, the low-polarized contribution (see Figure \ref{fig:local_not_in_mue.}) from the highest local emission cosines $\mu_\mathrm{e}$ can be significant in the total signal.

From Figure \ref{fig:K3_R_unpol_pdeg.} one may infer that higher spin produces lower overall amplitude of the reflected polarization for most of the configurations. This is because in the parts closest to the ISCO (which descends towards the BH for faster rotating BHs) there is more dramatic change of the polarization angle and the emission angle, leading to the globally depolarized signal. Figures \ref{fig:fig9_tot_neutral.} and \ref{fig:fig9_tot_ionized.} show the more subtle role of spin for the total spectra at infinity in various cases, when the primary is added. The further we reach out with the lamp-post height, the less of an impact spin has.
In the most inclined sources with the lowest lamp-post heights (i.e. when vast majority of studied geodesics are prone to large spacetime curvature), the distinction between the two studied spin cases is the clearest.

We expect the total polarization to decrease with height, since the fraction of unpolarized primary radiation increases due to light-bending effects. We note that for heights close to $h = 100 \textrm{ } GM/c^2$ we lack a fraction of the reflected light due to the default disc truncation at $r_{\mathrm{out}} = 400 \textrm{ } GM/c^2$. The top panels of Figures \ref{fig:fig9_tot_neutral.} and \ref{fig:fig9_tot_ionized.} remain qualitatively the same, if we extend the outer disc radius to $r_{\mathrm{out}} = 1000 \textrm{ } GM/c^2$, only with a slightly higher polarized fraction at the high lamp-post height tail. This monotonous behaviour of polarization with height is only distorted for lamp-posts closest to the BH, where the geometry of scattering varies a lot in the inner accretion disc and the strong-gravity effects are largest.

\subsection{Polarization angle}

\begin{figure*}
	\includegraphics[width=2.\columnwidth]{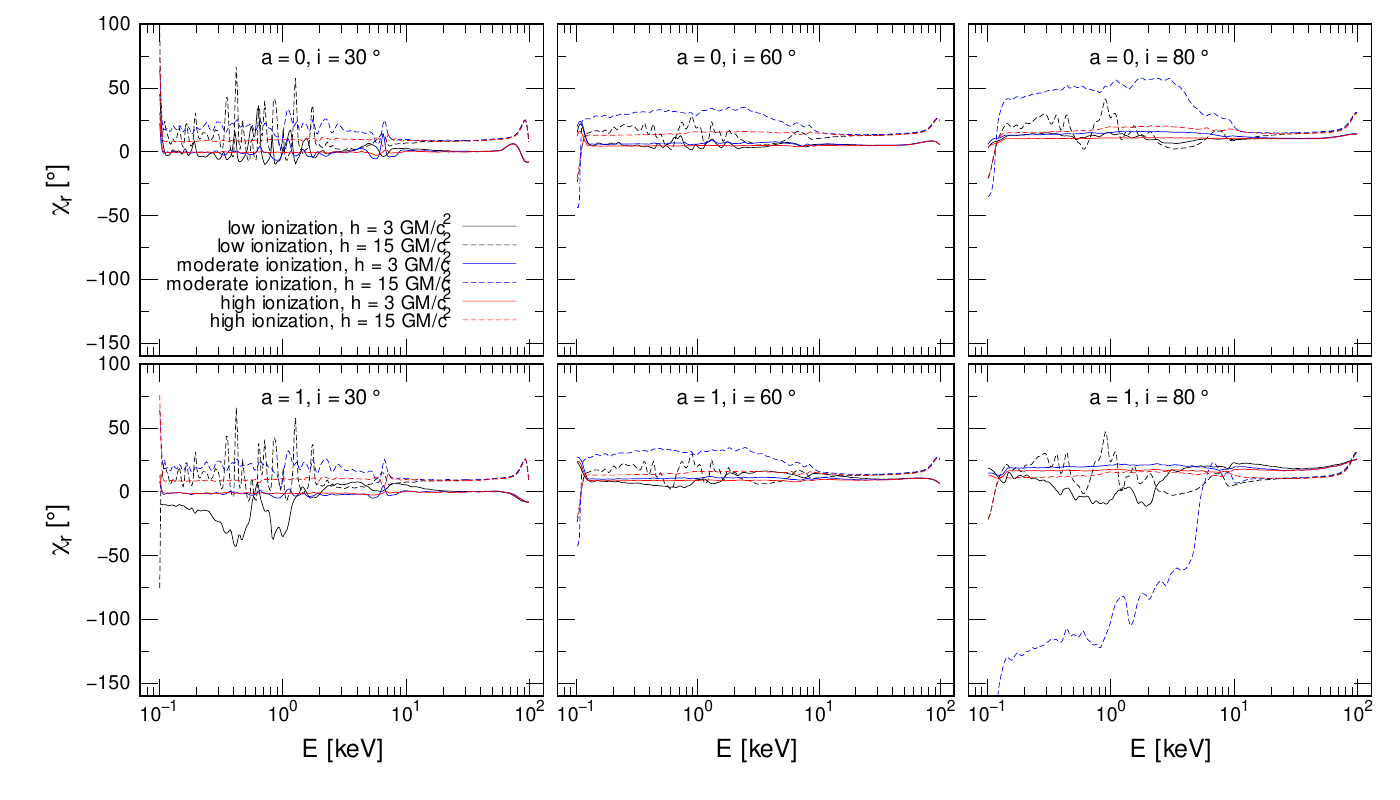}
	\caption{The reflected-only polarization angle, $\chi_\mathrm{r}$, versus energy for the same parametric setup as in Figure \ref{fig:K3_R_unpol_I.}, displayed in the same manner.}
	\label{fig:K3_R_unpol_pang.}
\end{figure*}

As in the previous sections, also in this one it is assumed that the primary radiation is unpolarized. This is why the total polarization angle $\chi$ is the same as the reflected-only polarization angle $\chi_\mathrm{r}$ at infinity.
Despite the fact that the polarization in this case can only originate in the disc re-processing, we note that the polarization angle is in general a very sensitive quantity to the general relativistic effects, as its rotation in the polarization plane can be significant even at large distances far from the black hole \citep[see figure 3 in ][]{Dovciak2008}.

The polarization angle energy dependence of the reflected radiation for the same parameter values as in Figures \ref{fig:K3_R_unpol_I.}, \ref{fig:K3_RP_unpol_I.}, \ref{fig:K3_R_unpol_pdeg.} and \ref{fig:K3_RP_unpol_pdeg.} is shown in Figure \ref{fig:K3_R_unpol_pang.}. One can see polarizaion angle variations due to unpolarized spectral lines at energies below $3\,$ keV and close to iron line complex that were already present in the local tables [see \cite{Podgorny2021} and Section \ref{local}]. In the same manner with the ionization parameter, $\xi$, in the local co-moving frame, the spectral lines are most dominant for the case of neutral disc (black lines). For high inclination, low height and high BH spin cases, we again observe the relativistic weighted integration and line broadening that smear and suppress these sharp energy features originating from the local reflection tables. 

Note that in the local reflection tables the continuum polarization angle is almost constant in energy with its value close to $\chi_\mathrm{r} \approx 90^\circ$, i.e. close to be parallel with the disc, if uniformly integrated in the angular space \citep{Podgorny2021}. When computing the relativistic reflection from the full accretion disc, the total polarisation angle comes out to be $ 0^\circ \lesssim \chi_\mathrm{r} \lesssim 20^\circ$ slightly increasing with height and we notice steeper increase for the heights above $\approx 40\,GM/c^2$ and non-constancy in energy (most significantly for higher $i$ and $a$). We account this to the interplay of distorted geometry of scattering (i.e. the selection of incident and emission angles -- including aberration effects -- that causes weighting of directions in the course of the relativistic integration), light-bending effects along the geodesics (from the disc to the observer only in this case of unpolarized primary) and the general-relativistic rotation of the polarization angle along the geodesics (from the disc to the observer only in this case of unpolarized primary), which are all more likely for increasing $a$ and $i$ and decreasing $h$.

\begin{figure*}
	\includegraphics[width=2.\columnwidth]{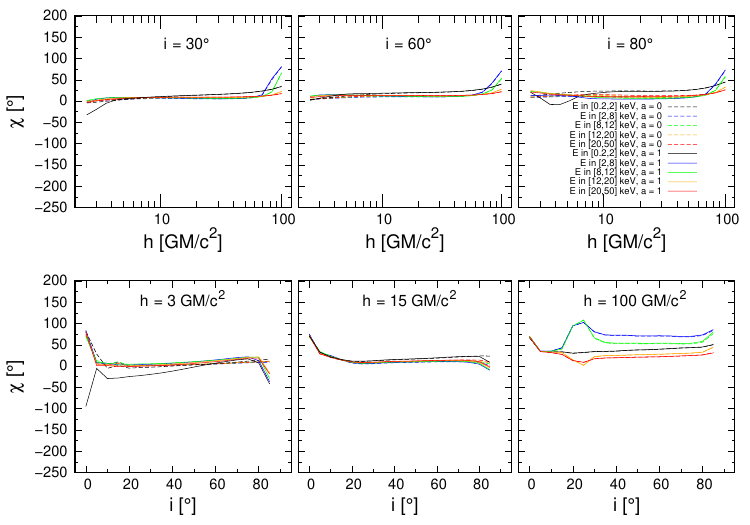}
	\caption{Top panel: the total average polarization angle, $\chi$, versus height of the primary point-source above the disc for disc inclinations $i = 30^{\circ}$ (left), $i = 60^{\circ}$ (middle) and  $i = 80^{\circ}$ (right), $\Gamma = 2$, unpolarized primary radiation, $M_{\textrm{BH}} = 1\times 10^8\,M_{\odot}$ and observed 2--10 keV flux $L_{\textrm{X}}/L_{\textrm{Edd}} = 0.001$, i.e. neutral disc. We show cases of two black-hole spins $a = 0$ (dashed) and $a = 1$ (solid) and for energy bands $E \in [0.2, 2]$ keV (black),  $E \in [2, 8]$ keV (blue), $E \in [8, 12]$ keV (green), $E \in [12, 20]$ keV (orange), and $E \in [20, 50]$ keV (red). Bottom panel: the total average polarization angle, $\chi$, versus disc inclination for heights of the primary point-source above the disc $h = 3 \textrm{ } GM/c^2$ (left), $h = 15 \textrm{ } GM/c^2$ (middle) and  $h = 100 \textrm{ } GM/c^2$ (right), for the same configuration as on the top panel.}
	\label{fig:fig10_tot_neutral.}
\end{figure*}
\begin{figure*}
	\includegraphics[width=2.\columnwidth]{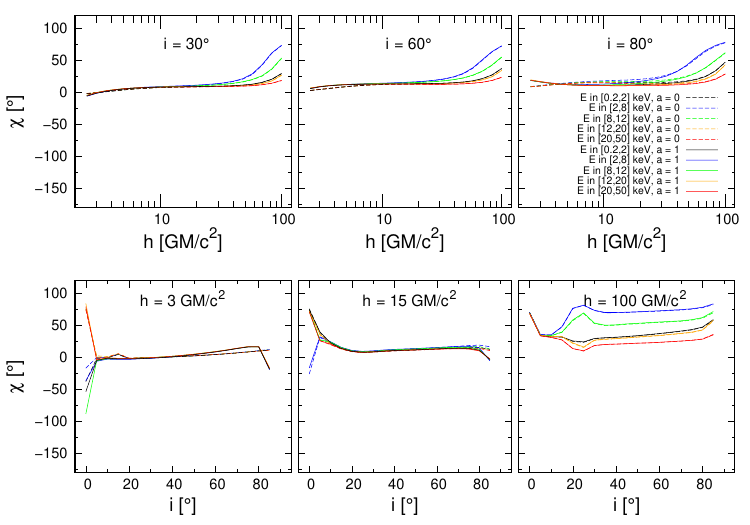}
	\caption{The total average polarization angle, $\chi$, with energy for ionized disc ($M_{\textrm{BH}} = 1\times 10^5\,M_{\odot}$ and observed 2--10 keV flux $L_{\textrm{X}}/L_{\textrm{Edd}} = 0.1$), otherwise for the same parametric setup as in Figure \ref{fig:fig10_tot_neutral.}, displayed in the same manner.}
	\label{fig:fig10_tot_ionized.}
\end{figure*}
We also provide the dependency on height and inclination of the average polarization angle in different energy bands and the two extreme spin cases. Figures \ref{fig:fig10_tot_neutral.} and \ref{fig:fig10_tot_ionized.} represent the neutral and ionized disc cases, respectively. Especially in the ionized case (or at higher energies in general), which are not distorted by spectral lines variations, we may inspect the continuum $\chi_\mathrm{r}$ dependency on height and inclination in a greater detail. The trend changes for low heights and high inclinations are caused by the relativistic effects. The height of the lamp-post only affects the different geometry of scattering here, as there is no polarization angle to be studied from the lamp to the disc, nor directly from the lamp to the observer. The polarization angle can take arbitrary value, if the polarization degree is approaching zero. Another situation when the polarization angle is arbitrary is when the photon direction is parallel with the normal to the disc, which defines the angle by projection to the polarization plane. The departures at high heights for the polarization angle can be explained by the first case (see Figures \ref{fig:fig9_tot_neutral.} and \ref{fig:fig9_tot_ionized.}), the departures at low inclinations by the second case in addition to the first. One can also notice on the dispersion in the color code (most prominently for the highest lamp-posts) the slight dependency on energy for the continuum $\chi_\mathrm{r}$ at infinity. 

\subsection{Non-zero incident polarization impact}\label{incident_pol}

\begin{figure}
	\includegraphics[width=1.\columnwidth]{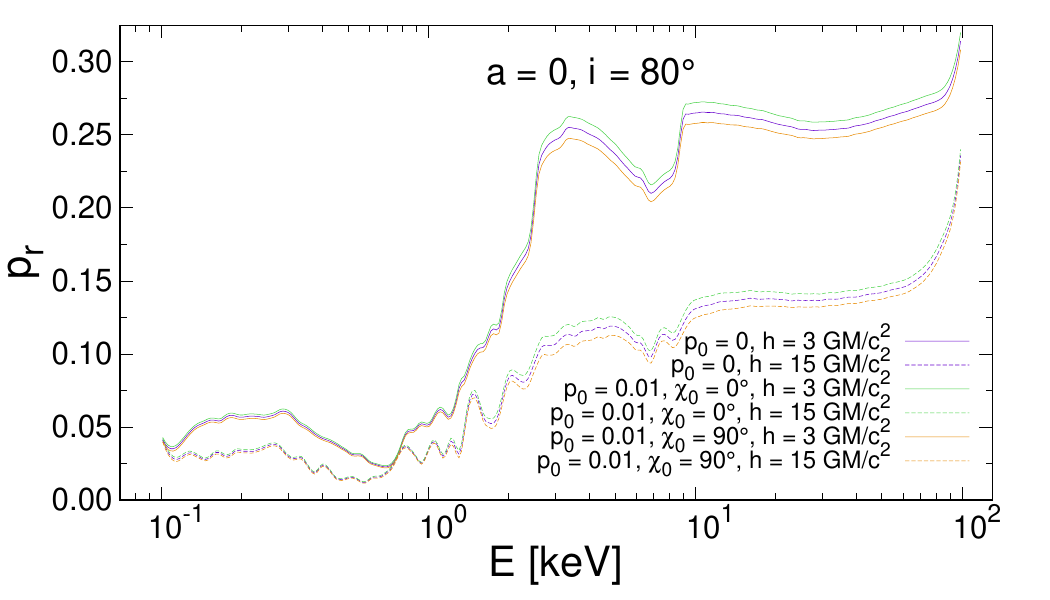}
	\caption{The reflected-only polarization degree, $p_\mathrm{r}$, versus energy of the accretion disc for distant observer, obtained by {\tt KYNSTOKES} for black-hole spin $a = 0$, disc inclination $i = 80^{\circ}$, $\Gamma = 2$ and for the neutral disc ($M_{\textrm{BH}} = 1\times 10^8\,M_{\odot}$ and observed 2--10 keV flux $L_{\textrm{X}}/L_{\textrm{Edd}} = 0.001$), using the {\tt STOKES} local reflection model in the lamp-post scheme. We show cases of two different heights of the primary point-source above the disc $h = 3 \textrm{ } GM/c^2$ (solid lines) and $h = 15 \textrm{ } GM/c^2$ (dashed lines), and three different polarization states of the primary source: $p_0 = 0$ (purple), $p_0 = 0.01$ and $\chi_0 = 0^{\circ}$ (green), $p_0 = 0.01$ and $\chi_0 = 90^{\circ}$ (orange).}
	\label{fig:K3A_R_polar_n_pdeg_unique.}
\end{figure}
\begin{figure}
	\includegraphics[width=1.\columnwidth]{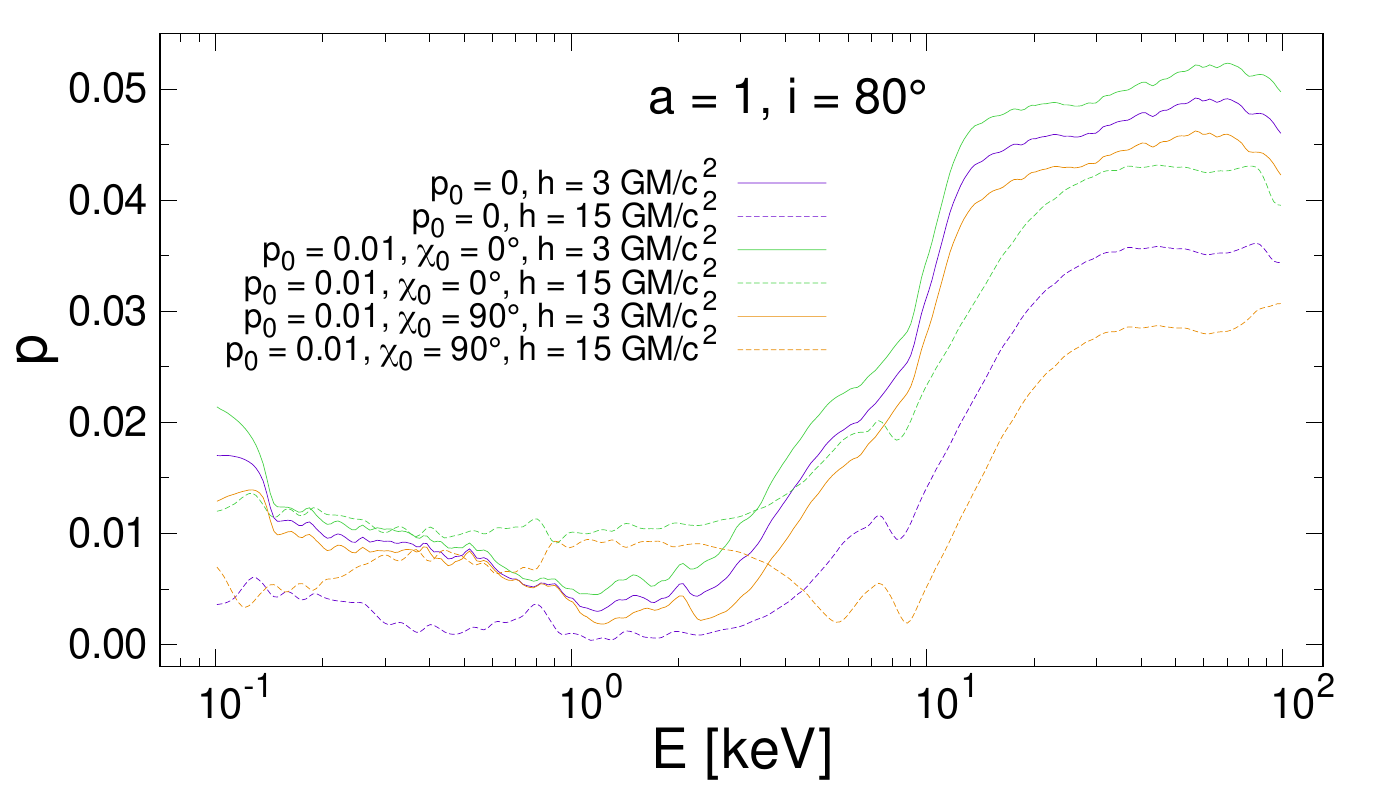}
	\caption{The total polarization degree, $p$, versus energy for black-hole spin $a = 1$, otherwise for the same parametric setup as in Figure \ref{fig:K3A_R_polar_n_pdeg_unique.}, displayed in the same manner.}
	\label{fig:K3A_RP_polar_n_pdeg_unique.}
\end{figure}
\begin{figure}
	\includegraphics[width=1.\columnwidth]{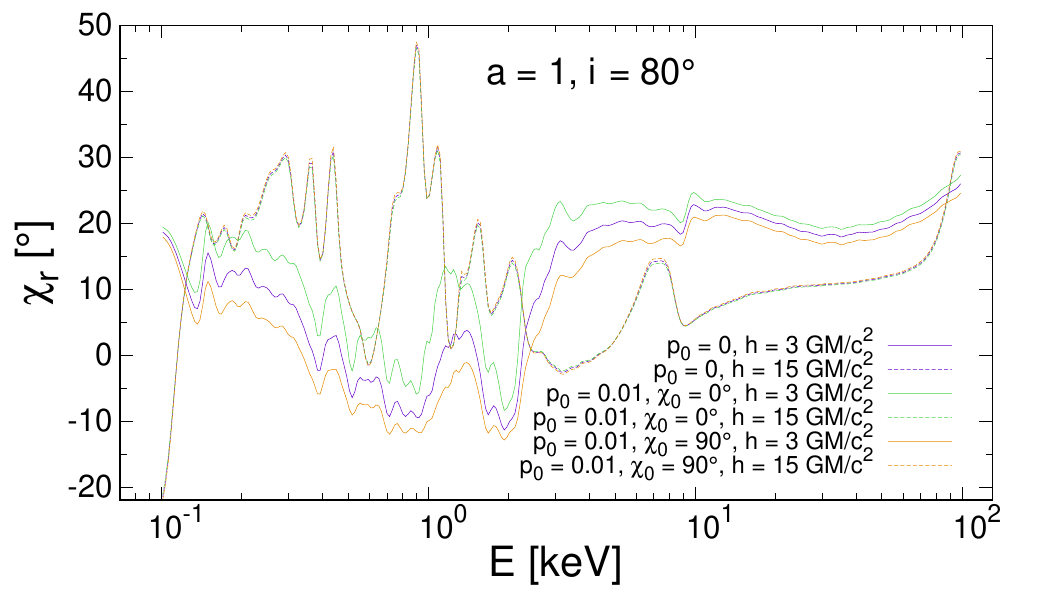}
	\caption{The reflected-only polarization angle, $\chi_\mathrm{r}$, versus energy for black-hole spin $a = 1$, otherwise for the same parametric setup as in Figure \ref{fig:K3A_R_polar_n_pdeg_unique.}, displayed in the same manner.}
	\label{fig:K3A_R_polar_n_pang_unique.}
\end{figure}
\begin{figure*}
	\includegraphics[width=2.\columnwidth]{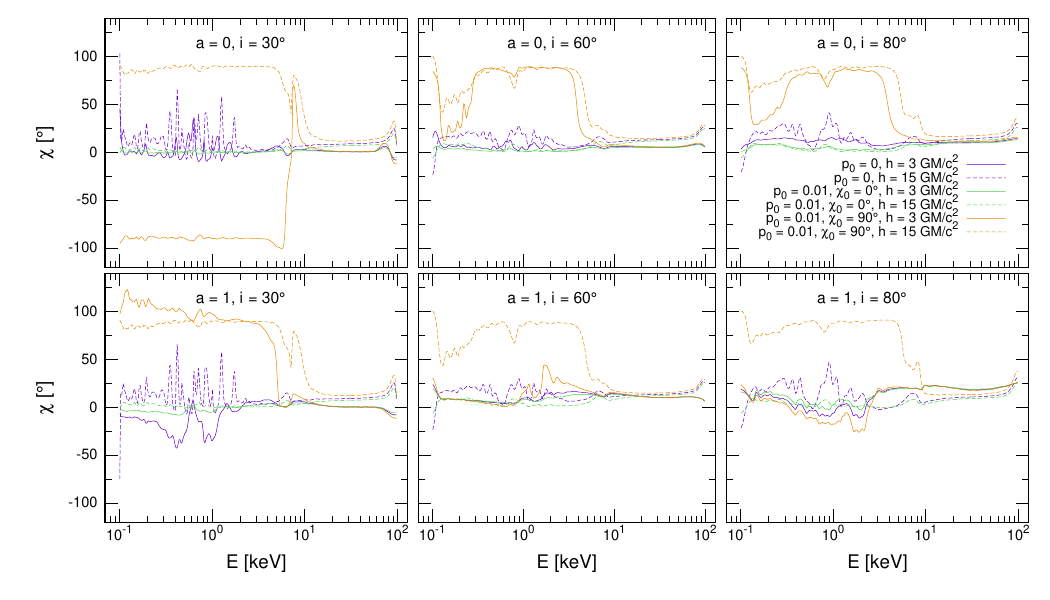}
	\caption{The total polarization angle, $\chi$, versus energy of the accretion disc for distant observer, obtained by {\tt KYNSTOKES} for black-hole spins $a = 0$ (top) and $a = 1$ (bottom), disc inclinations $i = 30^{\circ}$ (left), $i = 60^{\circ}$ (middle) and  $i = 80^{\circ}$ (right), $\Gamma = 2$ and neutral disc ($M_{\textrm{BH}} = 1\times 10^8\,M_{\odot}$ and observed 2--10 keV flux $L_{\textrm{X}}/L_{\textrm{Edd}} = 0.001$), using the {\tt STOKES} local reflection model in the lamp-post scheme. We show cases of two different heights of the primary point-source above the disc $h = 3 \textrm{ } GM/c^2$ (solid lines) and $h = 15 \textrm{ } GM/c^2$ (dashed lines), and three different polarization states of the primary source: $p_0 = 0$ (purple), $p_0 = 0.01$ and $\chi_0 = 0^{\circ}$ (green), $p_0 = 0.01$ and $\chi_0 = 90^{\circ}$ (orange).}
	\label{fig:K3A_RP_polar_n_pang_mainbody.}
\end{figure*}
\begin{figure*}
	\includegraphics[width=1.9\columnwidth]{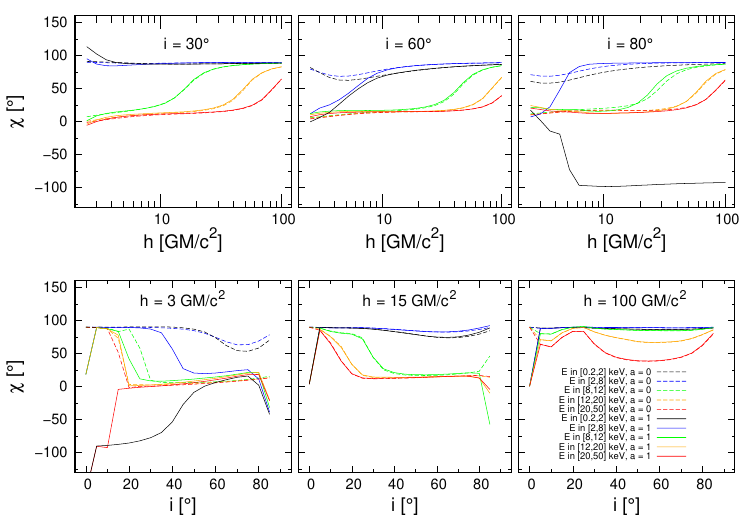}
	\caption{The total average polarization angle, $\chi$, with energy for primary polarization of $p_0 = 0.01$ and $\chi_0 = 90^{\circ}$, otherwise for the same parametric setup as in Figure \ref{fig:fig10_tot_neutral.}, displayed in the same manner.}
	\label{fig:fig10_tot_neutral_P2.}
\end{figure*}
In this section, we plan to discuss the impact of variable incident polarization that we studied in semi-realistic conditions of $p_0 = 0$ (purple lines in figures), $p_0 = 0.01$ and $\chi_0 = 0^{\circ}$ (green lines in figures), $p_0 = 0.01$ and $\chi_0 = 90^{\circ}$ (orange lines in figures) \citep{Beheshtipour2017, Tamborra2018, BeheshtipourThesis}. But we note that {\tt KYNSTOKES} allows for studies of arbitrary primary polarization state, which is calculated at the disc from the 0\% and 100\% polarized local reflection tables through (\ref{Sincident}). We preserve the distinction in height $h$ by the solid ($h = 3 \textrm{ } GM/c^2$) and dashed lines ($h = 15 \textrm{ } GM/c^2$) in the figures.
We did not find any significant impact on the spectral results for a distant observer \citep[a result already predicted by the local studies in][]{Podgorny2021}, therefore we reduce the discussion only to the polarization quantities.

Apart from the total polarization angle $\chi$ for the case of neutral disc (Figure \ref{fig:K3A_RP_polar_n_pang_mainbody.}), where the discussion is more interesting and the impact is large, the effect of non-zero incident polarization is qualitatively similar for the $a$ and $i$ cases studied. Therefore, Figures \ref{fig:K3A_R_polar_n_pdeg_unique.}, \ref{fig:K3A_RP_polar_n_pdeg_unique.} and \ref{fig:K3A_R_polar_n_pang_unique.} provide the resulting $p_r$, $p$ and $\chi_r$, respectively, with energy for one combination of $a$ and $i$ and the neutral disc. We also omit the results for highly ionized discs, as the stronger reflection contribution in such cases is responsible for the same primary polarization effects seen in the neutral reflection without the more interesting high contribution of primary radiation of particular polarization.

The polarization degree (total or reflected-only) tends to have qualitatively the same response to the tested incident polarization cases at hard X-rays for all other parameter values and it results in the change of the polarization degree of $\lesssim 1\%$. The originally horizontally polarized light (orange lines) is slightly diminished with respect to the unpolarized light (purple lines), while the vertically polarized light (green lines) is slightly enhanced. This effect is also seen if we test larger $p_0 = 2\%$ with linearly higher difference from $p_0 = 0\%$ on the output. Regarding the disc re-processing stage, in the local co-moving frame we already showed in \cite{Podgorny2021} that the $Q$ and $U$ Stokes parameters depend on the initial polarization state. This dependency is further complicated for a distant observer due to the superposition of photons with different polarization angles, which change along each geodesic differently. Now for the orange and green lines the relativistic rotation also matters for the photon trajectories between the lamp and the disc and between the lamp and the observer. In general the impact of this variable incident polarization of $p\leq 1\%$ on the observed polarization angle is $\lesssim 1.5^\circ$ in the reflected (continuum) component and $\lesssim 4^\circ$ when the primary is added, because it adds an unprocessed signal with constant direction of polarization, which disrupts the symmetry for a distant observer. Otherwise the results in both emergent polarization degree and angle are symmetric with respect to the unpolarized primary case when switching from horizontal to vertical incident polarization angle and keeping the same non-zero incident polarization degree.

The primary emission at infinity also disrupts the general pattern for the neutral disc case in the orange lines at $E<8\,$keV, i.e. the energy band, where the IXPE and eXTP instruments operate. This occurs for both, the total polarization degree and angle, and is due to the lack of reflected component at soft X-rays for the low-ionized disc. The total polarization degree for the neutral disc case tends to unite in the green and orange at the soft X-rays, because the primary radiation prevails over the weak reflected component and the total polarization degree obtains more its original $1\%$ value even in the orange lines (with horizontally oriented polarization at the lamp-post). It is the orange lines which have significantly different orientation of the polarization vector with respect to the dominant $\approx 0^\circ$ value for the originally unpolarized light (purple lines) after the disc re-processing and relativistic integration. Therefore this effect is the clearest in Figure \ref{fig:K3A_RP_polar_n_pang_mainbody.} for the polarization angle. Figure \ref{fig:fig10_tot_neutral_P2.} provides an equivalent of Figure \ref{fig:fig10_tot_neutral.} for the neutral disc case of the total $\chi$, but for $p_0 = 0.01$ and $\chi_0 = 90^\circ$. We show this figure to probe the dependency on energy, $h$, $i$ and $a$ of this effect visible in Figure \ref{fig:K3A_RP_polar_n_pang_mainbody.} in the orange lines from a different perspective. We may notice that the lowest energy bands tend to obtain the original $\chi_0 = 90^\circ$ value (or close by, since the rotation of the polarization angle from the lamp to the observer is negligible) and that the energy limit where this plays a role moves upwards for higher heights, as the reflected radiation looses its impact at infinity.

\subsection{Dependence on the photon index $\mathbf{\Gamma}$}

Let us provide a few comments to the dependency of all Stokes parameters for the unpolarized primary radiation on the $\Gamma$ parameter, which we held fixed until now. We did not discuss these dependencies in detail in \cite{Podgorny2021} on the local level, because the reflection tables produced by {\tt STOKES} had uneven grid in the $\xi$ parameter with respect to the $\Gamma$ parameter for technical reasons. Now that {\tt KYNSTOKES} performs all the necessary interpolations for us, we can reopen the discussion for a distant observer. We omit from the discussion the basic slope change in the total output for both the primary and reflected components, which is of course present. In general, this parameter also affects the disc ionization, which is theoretically more interesting. This is because if we fix the $L_{\textrm{X}}/L_{\textrm{Edd}}$, $M_{\textrm{BH}}$, and $h$ parameters, which mostly affect the ionization in the global setup, the definition (\ref{xi}) of $\xi$ imposes integration over the total energy range, and thus the photon index of the primary power-law is, along with the constant $n_\mathrm{H}$ assumption, the only remaining parameter that changes the local $\xi$.
\begin{figure*}
	\includegraphics[width=1.9\columnwidth]{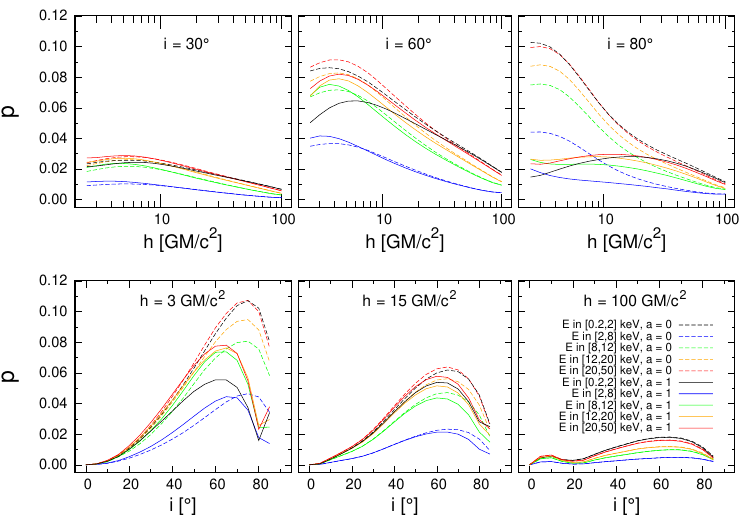}
	\caption{The total average polarization degree, $p$, with energy for $\Gamma = 3.0$, otherwise for the same parametric setup as in Figure \ref{fig:fig9_tot_ionized.}, displayed in the same manner.}
	\label{fig:fig9_tot_ionized_30.}
\end{figure*}

What we indeed observe in our reflected-only and total spectral results for various $\Gamma$ tested is that higher photon index makes the spectral lines and absorption originating in the local disc re-processing more pronounced, since less energetic photons ionize the disc (see e.g. the local spectral behaviour with $\xi$ integrated in angular space in Figure \ref{fig:local_not_in_mue.}). This is consistent with the local reflection dependencies on $\Gamma$ provided in figure 6 in \cite{Garcia2013}, which discusses the {\tt XILLVER} tables obtained by a completely different method. This brings additional confidence to the use of our new {\tt STOKES} local tables in addition to the discussion in \cite{Podgorny2021}, including their local polarization component, which cannot be compared to any similar simulation in the current literature.

Since line features and absorption caused by disc ionization are also imprinted to the polarization degree with energy for a distant observer (see Figures \ref{fig:K3_R_unpol_pdeg.} and \ref{fig:K3_RP_unpol_pdeg.}), $\Gamma$ affects the energy dependence of polarization degree at infinity induced by reflection. With higher $\Gamma$, we tend to see sudden depolarization between 0.8--4 keV, where the polarization degree with energy had a nearly flat curve in the moderately ionized (blue curves) and highly ionized (red curves) cases in Figures \ref{fig:K3_R_unpol_pdeg.} and \ref{fig:K3_RP_unpol_pdeg.}. We document the resulting dependency of the total polarization degree on $\Gamma$ in different energy bands by providing a remake of Figure \ref{fig:fig9_tot_ionized.} (showing the results for $\Gamma = 2.0$) but for $\Gamma = 3.0$ in Figure \ref{fig:fig9_tot_ionized_30.}. Notice the extended $y$-axis and the blue lines (representing $E \in [2, 8]$ keV, which is the energy range of the IXPE and eXTP polarimeters) being significantly lower than rest of the curves. It further illustrates how the amount of reflected flux and how the disc ionization are connected to the $h$, $a$, $i$ parameters, holding the $M_{\textrm{BH}} = 1\times 10^5\,M_{\odot}$ and observed 2--10 keV flux $L_{\textrm{X}}/L_{\textrm{Edd}} = 0.1$ fixed with respect to the previously shown highly ionized disc case with $\Gamma = 2.0$ in Figure \ref{fig:fig9_tot_ionized.}. The polarization angle for a distant observer is also affected by $\Gamma$ in terms of enhanced or suppressed presence of the line features.

\section{Discussion}\label{discussion}

\subsection{Comparison with other computations}

We refer to \cite{Podgorny2021} where we already compared the spectral part of the {\tt STOKES} local tables to the {\tt XILLVER} \citep{Garcia2010, Garcia2011, Garcia2013} and {\tt REFLIONX} \citep{Ross1993, Ross1999, Ross2005} tables directly in the local co-moving frame and discussed possible reasons for discrepancies. Let us now compare only the polarization part of the output in the same lamp-post integration scheme, but with different local reflection computations. From the same {\tt KY} package \citep{Dovciak2004b, Dovciak2011} we will use the {\tt KYNLPCR} routine, which uses the {\tt NOAR} spectral reflection computations \citep{Rozanska2002} and the commonly used Chandrasekhar's analytical approximation \citep{Chandrasekhar1960} of single-scattering induced polarization. We will provide the results for the latest version of {\tt KYNLPCR}. Therefore, our results may differ from the previous publications using {\tt KYNLPCR} \citep{Dovciak2004b,Dovciak2011}, because of the latest updates mentioned in Section \ref{models}. By performing the identical relativistic integration (see Section \ref{global}) and assuming a neutral disc, we aim to show the difference and importance of the local reflection tables.

For simplicity, we will focus on the $h = 3 \textrm{ } GM/c^2$ and $\Gamma = 2$ case only, but let us keep the display scheme of spins $a = 0$ (top) and $a = 1$ (bottom) and disc inclinations $i = 30^{\circ}$ (left), $i = 60^{\circ}$ (middle) and  $i = 80^{\circ}$ (right). We truncate the energy range below $1\,$keV for reasonable comparison due to the neutrality of the {\tt NOAR} computations, which are used in {\tt KYNLPCR} and which thus lack a soft X-ray component. We do not show the ionized case comparison for the same reason and only show neutral disc in {\tt KYNSTOKES} ($M_{\textrm{BH}} = 1\times 10^8\,M_{\odot}$ and observed 2--10 keV flux $L_{\textrm{X}}/L_{\textrm{Edd}} = 0.001$). We keep the color code from Section \ref{results} for different incident polarization set by the $p_0$ and $\chi_0$ values. We will use solid lines for the {\tt KYNSTOKES} results and dashed lines for the {\tt KYNLPCR} results.

\begin{figure*}
	\includegraphics[width=2.\columnwidth]{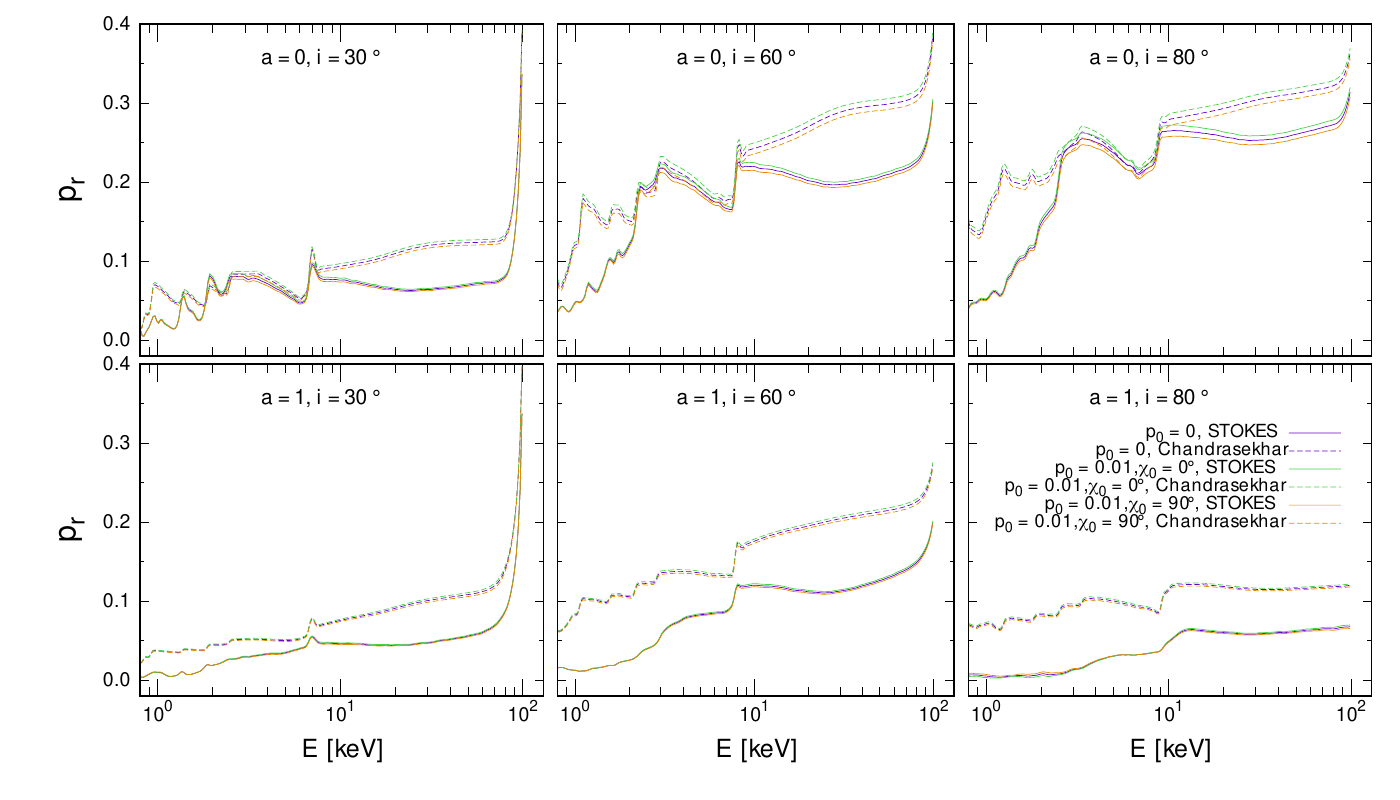}
	\caption{The reflected-only polarization degree versus energy, $p_\mathrm{r}$, of the accretion disc for distant observer in the lamp-post scheme for black-hole spins $a = 0$ (top) and $a = 1$ (bottom), disc inclinations $i = 30^{\circ}$ (left), $i = 60^{\circ}$ (middle) and  $i = 80^{\circ}$ (right), $\Gamma = 2$, height of the primary point-source above the disc $h = 3 \textrm{ } GM/c^2$ and neutral disc (for $M_{\textrm{BH}} = 1\times 10^8\,M_{\odot}$ and observed 2--10 keV flux $L_{\textrm{X}}/L_{\textrm{Edd}} = 0.001$). We compare cases of {\tt KY} disc integration with the {\tt STOKES} local reflection tables (solid lines) and {\tt NOAR} local reflection tables with Chandrasekhar's analytical approximation for polarization (dashed lines), and three different polarization states of the primary source: $p_0 = 0$ (purple), $p_0 = 0.01$ and $\chi_0 = 0^{\circ}$ (green), $p_0 = 0.01$ and $\chi_0 = 90^{\circ}$ (orange).}
	\label{fig:K5_p_Ln.}
\end{figure*}
\begin{figure*}
	\includegraphics[width=2.\columnwidth]{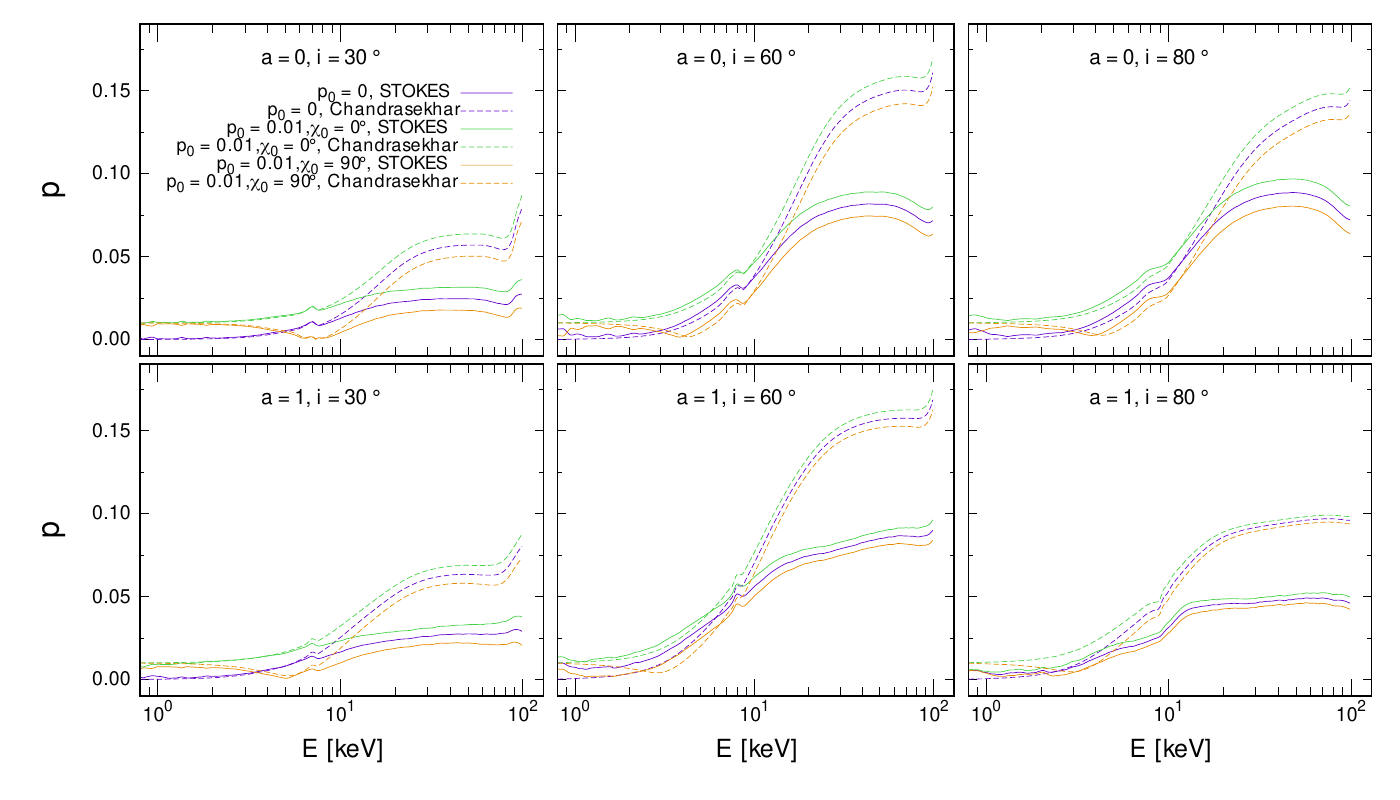}
	\caption{The total polarization degree, $p$, versus energy for distant observer for the same parametric setup as in Figure \ref{fig:K5_p_Ln.}, displayed in the same manner.}
	\label{fig:K5_p_Ln_RP.}
\end{figure*}
\begin{figure*}
	\includegraphics[width=2.\columnwidth]{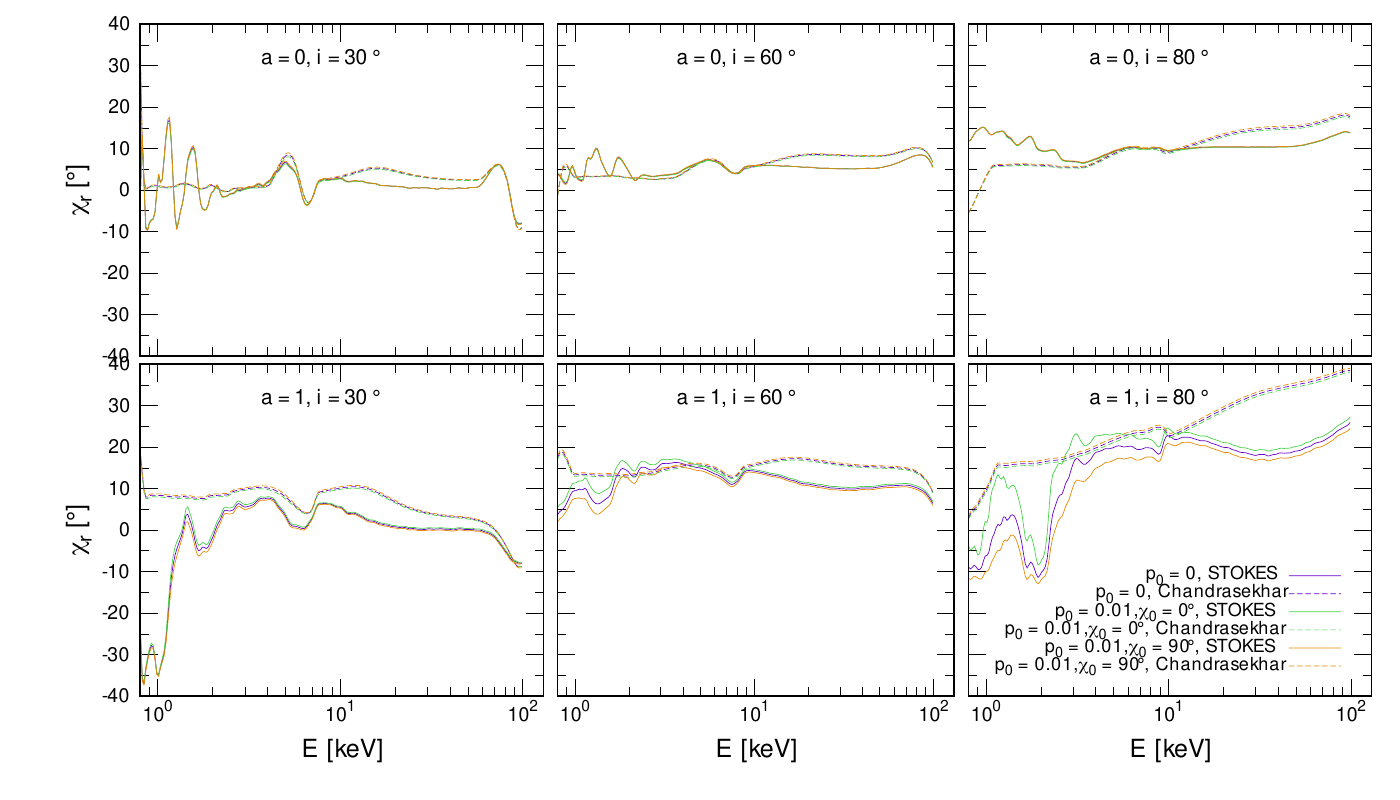}
	\caption{The reflected-only polarization angle, $\chi_\mathrm{r}$, versus energy for distant observer for the same parametric setup as in Figure \ref{fig:K5_p_Ln.}, displayed in the same manner.}
	\label{fig:K5_Psi_Ln.}
\end{figure*}
\begin{figure*}
	\includegraphics[width=2.\columnwidth]{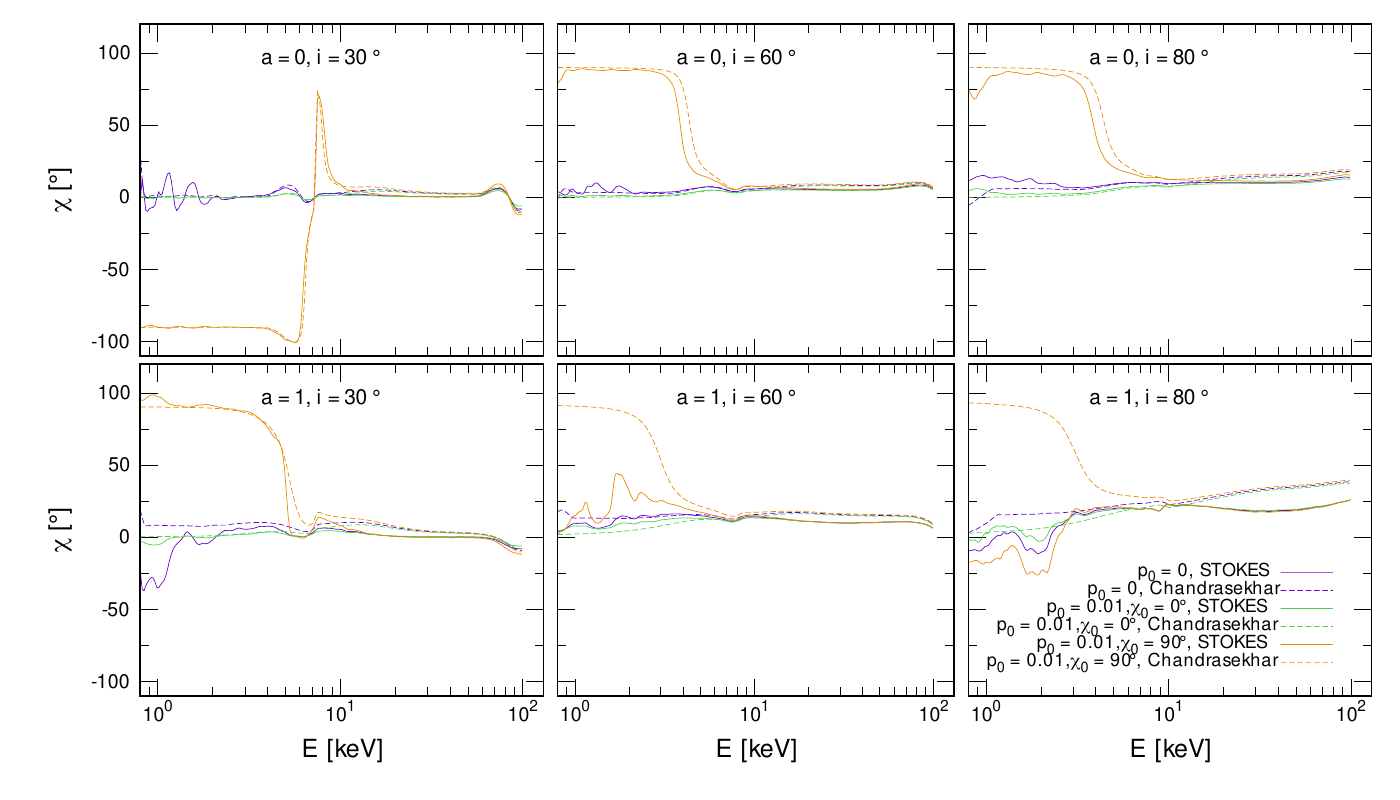}
	\caption{The total polarization angle, $\chi$, versus energy for distant observer for the same parametric setup as in Figure \ref{fig:K5_p_Ln.}, displayed in the same manner.}
	\label{fig:K5_Psi_Ln_RP.}
\end{figure*}
Figures \ref{fig:K5_p_Ln.} and \ref{fig:K5_p_Ln_RP.} show the reflected-only and total polarization degree with energy. Figures \ref{fig:K5_Psi_Ln.} and \ref{fig:K5_Psi_Ln_RP.} show the same comparison for the reflected-only and total polarization angle with energy. Despite a general similarity of the two simulated polarization degree shapes with energy for a distant observer, in all of the displayed configurations in Figures \ref{fig:K5_p_Ln.} and \ref{fig:K5_p_Ln_RP.}, we observe a significant depolarization in {\tt KYNSTOKES} results with respect to the predictions by Chandrasekhar's approximation in the Compton hump region above $10\,$keV, where only continuum processes appear. This is the energy band where e.g. the XL-Calibur polarimeter operates. The result was already predicted by direct comparison on the local level in \cite{Podgorny2021}, and it appears even in the total emission detected at spatial infinity. The main reason for the depolarization by approximately 55\% is multiple scatterings, which are considered by the Monte Carlo simulator {\tt STOKES} \citep{Goosmann2007,Marin2012,Marin2015,Marin2018} and which cause larger dispersion in the emergent polarized signal. Also the elastic Rayleigh scattering assumed by the Chandrasekhar's approximation is less realistic in such physical environment when compared to the proper Compton down-scattering taken into account by {\tt STOKES}.

The unification of green and orange lines for the total polarization degree $p$ in Figure \ref{fig:K5_p_Ln_RP.} towards the mid and soft X-rays for {\tt KYNLPCR} (dashed lines) now confirms our previous explanation of this feature for {\tt KYNSTOKES} (solid lines) in Section \ref{incident_pol}. There is a complete lack of reflected component at soft X-rays for a distant observer in the {\tt KYNLPCR} code that assumes a cold neutral disc. A similar (but non-zero) weak reflected component with respect to the primary radiation is present at mid and soft X-rays in the {\tt KYNSTOKES} computations for the neutral disc case, which we discuss here. Thus, the polarization degree is in the case of orange lines prone to obtain its original 1\% value given by the dominant direct primary radiation.

This is even more clear on the display of total polarization angle in Figure \ref{fig:K5_Psi_Ln_RP.}, where the dashed orange lines obtained by {\tt KYNLPCR} also follow the original $\chi_0 = 90^\circ$ value below $\lesssim 8\,$keV. We see that it happens in more cases for the dashed orange lines than for the solid orange lines produced by {\tt KYNSTOKES}. This is due to the true neutral disc assumption and a complete lack of the reflected component at soft X-rays in {\tt KYNLPCR}. In Figures \ref{fig:K5_Psi_Ln.} and \ref{fig:K5_Psi_Ln_RP.} we observe some disunity between the dashed and solid lines in the region above $10\,$keV, which cannot be accounted to the treatment of spectral lines or reflectivity by the neutral or almost neutral disc, but is again due to the characteristic Compton recoil dominating at these energies, which cannot be imitated by the Chandrasekhar's analytical approximation. Nonetheless, the predictions for the reflected-only and total polarization angle by {\tt KYNLPCR} and {\tt KYNSTOKES} are also qualitatively similar for the tested parameters.

\subsection{Future modeling improvements}\label{future_improvements}

Our current model contains many simplifications and much room for improvement that we aim to achieve in the near future. Staying in the lamp-post scheme, which itself is a simplified concept \citep[see e.g.][]{Marinucci2018,Poutanen2018, Kubota2018}, we may first develop the non-isotropic lamp-post emission. This would certainly improve precision in polarization quantities, especially for the reflected component, because the disc is expected to receive variable flux and polarization, depending on each position and each incident angle. For the realistic primary flux and primary polarization estimates (even departing from the strict point-like source, assuming a small 3D patch on the rotational axis of the disc), we may use the {\tt MONK} code \citep{Zhang2019b}. This Monte Carlo code is able to simulate Comptonization of the disc thermal radiation in the hot corona with variable 3D shapes.

Concerning the disc re-processing computed by the current {\tt STOKES} Monte Carlo method, we already pointed out in \cite{Podgorny2021} a number of possibilities for improvement. It is desired to incorporate effects of Comptonization, synchrotron emission, or thermal disc radiation into the {\tt STOKES} simulations.
Inclusion of the thermal treatment and absorption in X-rays \citep[see][]{Taverna2020b} to this reflection study would allow to use the current model for X-ray binary accretion discs, if also higher disc densities ($10^{19}$--$10^{25} \ \textrm{cm}^{-3}$) and temperatures ($10^{5}$--$10^{7.5} \ \textrm{K}$) were considered \citep{Shakura1973, Novikov1973, Reynolds2003, Abramowicz2013, Compere2017, Kubota2018}. Once consistent local X-ray reflection tables for XRBs are produced, we may easily produce the global XRB model with the same {\tt KY} relativistic integration tools. This is timely and desirable, as higher fluxes are expected from the Galactic sources and photon-demanding X-ray polarimetry will give more insights into the inner accretion phenomena of stellar-mass black holes \citep[see recent observational results by ][]{Krawczynski2022c, Veledina2023, Podgorny2023, Rawat2023, Kushwaha2023, Ratheesh2023, Cavero2023}.

Regarding the accretion disc studies, it would be also beneficial to release the assumption of constant density in our model and allow more flexibility in elemental abundance and the shape of the primary X-ray radiation [including a possible selection of cut-offs that substantially influence the reflected output \citep{Garcia2013}]. The advantage of the {\tt STOKES} local spectral tables over the {\tt XILLVER} tables is the availability of the full ($\mu_\mathrm{i}$, $\Phi_\mathrm{e}$, $\mu_\mathrm{e}$) local angular space, while the {\tt XILLVER} tables are integrated in the $\mu_\mathrm{i}$ and $\Phi_\mathrm{e}$ parameters. The precision in angular behavior of the local geometry of scattering may become important for the relativistic integration over the disc in curved space, especially in the lamp-post coronal geometries and especially for the treatment of polarization, which is very sensitive to geometry of scattering. We stress that none of the local reflection tables tested in \cite{Podgorny2021} and this study assumed a stratified disc atmosphere in hydrostatic equilibrium. Refraining from the constant disc density assumption may lead to significantly different results in both continuum and emission line shapes; however, direct comparisons are not easy \citep{Rozanska2002, Podgorny2021} and any such attempt would be beneficial. Additionally, as already mentioned in Section \ref{local}, the resolution of the radiative transfer inside the disc photosphere might be failing to pick up strong resonance lines, thus underestimating the polarization, mostly below 2 keV.

Lastly, a significant simplification of the global model is the neglect of returning radiation, i.e. those rays that originate in the corona and are re-processed in the disc that would have such trajectories in curved spacetime emerging from the disc to be re-processed again in the disc, possibly multiple times before reaching the observer. This problem was addressed from the X-ray spectroscopic view recently e.g. in \cite{Dauser2022} for a lamp-post corona and from the X-ray polarimetric view in detail in \cite{Schnittman2009, Schnittman2010} but for extended coronal geometries and with a focus on thermally radiating XRBs. From \cite{Dauser2022} one can deduce that higher order re-processings affect the spectral shape due to contribution from inner regions of the disc ($\lesssim 10 \ r_\textrm{g}$), most importantly for high BH spins ($\gtrsim 0.9$) and low lamp-post heights ($\lesssim 5 \ r_\textrm{g}$) and highly ionized disc, resulting in higher unabsorbed fractions. The polarization arising from self-irradiation of either thermal emission or extended coronal emission reprocessed once from the disc, assuming the analytical formulae for diffuse reflection \citep{Chandrasekhar1960}, is azimuthally dependent, producing net fraction $\sim 10$\% with polarization angle parallel to the disc axis, both constant in energy \citep{Schnittman2009, Schnittman2010}.

The inner regions are however the most affected by the GR effects causing a non-trivial geometrical superposition of polarization, which makes the effect of returning radiation on polarization more unpredictable in our case. This work showed that the \textit{once} reflected X-rays are comparably or more significantly polarized compared to the thermal radiation arising from the disc in \cite{Schnittman2009} simulations, being the primary source of emission therein. The polarization of such thermal radiation can be approximated by an analytical prescription for scattering-induced polarization in semi-infinite atmospheres \citep{Chandrasekhar1960}, which results in monotonously rising polarization fraction with inclination from 0\% (viewed pole-on) up to about 12\% (viewed edge-on). Thus additional polarization due to returning radiation would perhaps impact our results similarly or less than being superimposed on the direct disc emission as shown in \cite{Schnittman2009}. However, the flux angular distribution will differ, as well as flux contributions onto and from different disc regions. Because the incident polarization state impacts the reflected polarization state to a considerable extent (see Section \ref{incident_pol}), it is also unclear whether we would see the same effects of returning radiation due to different polarization of the returning primary thermal radiation \cite{Schnittman2009} or once re-processed Comptonized power-law emerging from the equatorial plane (this work). These arguments of unknown impact of incident polarization and both incident and emergent flux angular distribution per each disc region on the returning radiation hold for comparisons with extended or patchy coronae alongated in the disc plane that are considered in \cite{Schnittman2010} and that (combined with the thermal radiation) represent yet another case from pure thermal emission, or a pure lamp-post model with prescribed coronal emission. This case is further complicated by interaction of the returning radiation with the coronal scattering regions again, making direct comparisons with the model presented in this paper misleading. It is mentioned though in \cite{Schnittman2010} that the wedge corona produces similar returning radiation image to the returning thermal radiation and equivalently a net vertical polarization, despite suppressing the returned flux from larger radii.
 
Therefore, by addition of returning radiation we would expect an increase of polarization results presented here that show on average the same vertical orientation of the polarization vector. The polarization orientation of our results and the returning radiation from \cite{Schnittman2009, Schnittman2010} is energy-independent and the rescattered contribution is more significant in the hard X-rays. Hence, an increase of polarization fraction by a few percent in the 10--100 keV band would be anticipated without the switch in polarization angle in 1--10 keV seen in \cite{Schnittman2009, Schnittman2010} due to a model with first order polarization orientation roughly parallel to the disc axis, not perpendicular. However, an implementation of self-irradiation to {\tt KYNSTOKES} would be necessary to investigate such effects in detail and prove the above predictions.

\section{Summary and Conclusions}\label{conclusion}

In this paper we discussed a new {\tt XSPEC} model {\tt KYNSTOKES} that produces a complete spectropolarimetric output for a distant observer of the primary and reflected X-ray emission from black-hole inner-accreting regions. We assumed the lamp-post coronal scheme with a hot compact source of X-ray primary power-law emission located on the rotational axis of the disc at some height above the BH. We made use of the newly computed local reflection tables \citep{Podgorny2021} by the codes {\tt TITAN} \citep{Dumont2003} and {\tt STOKES} \citep{Goosmann2007,Marin2012,Marin2015,Marin2018} that provide a 3D Monte Carlo simulation of the Stokes parameters emergent from a semi-infinite scattering atmosphere with constant density $n_\mathrm{H} = 10^{15}\,\textrm{cm}^{-3}$. Various line and continuum processes are incorporated, including the physics of absorption, re-emission and Compton down-scattering, which is expected to be a dominant source of polarization in the reflected X-rays, especially above $10\,$keV. The presented model is mainly intended for use for radio-quiet AGN inner-accretion studies. {\tt KYNSTOKES} performs the necessary interpolation of the local tables and integrates them over the geometrically thin accretion disc in the equatorial plane using all special- and general-relativistic effects. The energy-dependent Stokes parameters for a distant observer are then provided by {\tt KYNSTOKES} for a range of parameters, such as the spin of the BH, the disc inclination, the height of the primary source above the BH, the luminosity of the primary source, the mass of the BH, the primary power-law photon index, the primary polarization state, the disc emitting area, or possible partial obscuration.

Using these techniques, we first showed a new analysis of the reflected emission in the local co-moving frame complementary to \cite{Podgorny2021} performing a simple integration of the local tables in the angular space, which facilitates the discussion of the global results. Then we produced examples of the distant spectral and polarimetric output for different configurations. We choose different intrinsic luminosities and BH masses to study spectral and polarization properties in different disc ionization states. The most prominent reflection spectral features, such as the Fe K$\alpha$ line at $6$--$7 \textrm{ keV}$, a forest of lines below $3\,$keV, and the Compton hump above $10\,$keV, are heavily relativistically smeared. This global spectral behavior resembles well the predictions previously appearing in the literature, which provides confidence to the polarization results that are produced simultaneously by this numerical method, yet uniquely in the context of current research.

The polarimetric quantities produced by {\tt KYNSTOKES} represent a significant improvement over the results presented in \cite{Dovciak2011} that used similar integrating techniques and the Chandrasekhar's analytical approximation for local scattering \citep{Chandrasekhar1960}. For the case of unpolarized primary radiation, the reflected polarization angle tends to be oriented parallel with the disc axis of symmetry ($\approx 0^\circ$) at all studied energies for a distant observer. In the most favourable, yet realistic parametric configurations, the reflected polarized fraction can reach up to 25\% in the Compton hump region and up to 9\% in the same region when an upolarized primary radiation is included. We also studied the impact of non-zero primary polarization. The effect was negligible for the spectral results, while it was noticeable for the polarization degree and angle at different energy bands, especially in the configurations of nearly neutral disc when the direct primary radiation is included.

We also provided a comparison of the polarization results with the Chandrasekhar's analytical approximation for single scattering, previously used in the {\tt KY} codes \citep{Dovciak2011}. The newly considered multiple scatterings for the disc re-processing predict a decrease of the total polarization degree by approximately 55\% in the Compton hump region with respect to the previously used analytical approach within the same disc integration method.

We believe that the presented results on polarization by reflection in black-hole accretion discs will be useful for data fitting of real AGN and XRB sources observed by the X-ray polarimeters on board of current and future missions such as IXPE \citep{Weisskopf2022}, XL-Calibur \citep{Abarr2021}, or eXTP \citep{Zhang2016, Zhang2019}. The general-relativistic effects tend to modify the observable polarization quantities in a more radical way than the pure spectral profiles. The relativistic reflection creates a significant amount of polarized signal, which is heavily dependent on the system parameters. This raises the opportunities of constraining physical properties of such regions in the future and it will help to remove some parametric degeneracies known from the X-ray spectral studies.

\section*{Acknowledgements}

JP acknowledges financial support from the Charles University, project GA UK No. 174121, and from the Barrande Fellowship Programme of the Czech and French governments. MD and VK thank for the support from the Czech Science Foundation project GACR 21-06825X. JP, AR and VK thank the Czech-Polish mobility program (M\v{S}MT 8J20PL037 and PPN/BCZ/2019/1/00069) and the European Space Agency PRODEX project 4000132152. AR was supported by the Polish National Science Center grant No. 2021/41/B/ST9/04110. GM acknowledges financial support from Italian Space Agency (ASI-INAF-2017-12-H0). JP, MD and VK thank also the institutional support from the Astronomical Institute RVO:67985815.

\section*{Data Availability}\label{kynstokes_repository}

The model {\tt KYNSTOKES} underlying this article (created by authors of this article), including user instructions, is available at \url{https://projects.asu.cas.cz/dovciak/kynstokes}.



\bibliographystyle{mnras}
\bibliography{example} 




\appendix

\section{General-relativistic treatment of the polarization state}\label{calculations}

Let us provide how the GR calculations of the change of the polarization angle from the lamp to the observer and from the lamp to the disc are implemented in {\tt KYNSTOKES} in detail. We use similar approach as in \cite{Connors1977, Connors1980, Dovciak2004, Dovciak2004c,Dovciak2004d}, where more theoretical background is given. The polarization angle $\chi$ in the local reference frame is defined as increasing in the counter-clockwise direction in the incoming view of photon. It is the angle between the unit spatial polarization 3-vector $\Vec{f}$ and a disc normal projected to the polarization plane (perpendicular to the photon's momentum). Here, we first provide an overview of the general procedure and then show the derivation of the change of the polarization angle for the two cases of polarized radiative transfer that have been newly added to {\tt KYNSTOKES} in order to allow non-trivial polarization states of the coronal emission. We will use Greek subscript and superscript letters for denoting all 4-vector components and round brackets with latin letters for components in particular local reference frame (defined by a tetrade) as is usual.

We denote the local polarization vector that is defined in local reference frame
of the primary source on the axis as a unit spatial polarization 3-vector, $\vec{f_0}$. If the local reference frame is defined by an ortonormal tetrade $e_{\rm (a)}^\mu$, then the four components of the polarization 4-vector in this tetrade are $f_0^{\rm (a)}\equiv (0, \vec{f_0})$ and corresponding polarization 4-vector is $f_0^\mu = e_{\!\rm (a)}^\mu f_0^{\rm (a)}$.
The polarization 4-vector is then parallelly transported along a geodesic from the lamp either to a particular point at the disc in the equatorial plane, or to a particular direction at spatial infinity representing a distant observer. At the ending point of the parallel transport, the polarization 4-vector $f^\mu$ will, generally, possess a non-zero time component in the local reference frame. To keep all the properties of the polarization 3-vector after the transformation, we need to subtract a multiple of a null vector so that the time component is set to zero and the space part is not rotated. This means performing a transformation
\begin{equation}\label{reduction}
    f'^{\textrm{(a)}} = f^{\textrm{(a)}} - \frac{f^\textrm{(t)}}{p^\textrm{(t)}}p^{\textrm{(a)}}
\end{equation}
as the last step to obtain a properly defined polarization vector in the local reference frame at a particular disc point or for a particular distant observer.

As usual in the polarized radiative transfer in GR, to be efficient, we use the Walker-Penrose theorem \citep{Walker1970} proving the existence of a conserved complex quantity $K_\textrm{S} = \kappa_1 + i\kappa_2$ characterizing parallel transport along null geodesics in Kerr space-time. The Walker-Penrose constant, $K_\textrm{S}$, connects the 4-vector components of parallelly transported vector, $f^\mu$, to the photon momentum, $p^\mu$, as a tangent to given geodesic in the following way
\begin{equation}\label{KS}
\begin{split}
    K_\textrm{S} = \frac{1}{r+ia\cos{\theta}} &  \{(r^2+a^2)(p_rf_t-p_tf_r) \\
    &\quad +a(p_rf_{\varphi}-p_{\varphi}f_r) \\
    &\quad + \frac{i}{\sin{\theta}}[p_\theta f_\varphi - p_\varphi f_\theta \\
    &\quad - a\sin^2{\!\theta}\,(p_t f_\theta - p_\theta f_t) ] \}.
\end{split}
\end{equation}

In our case this constant characterizes parallel transport of a polarization 4-vector, $f^\mu$, along a light ray. This constant is uniquely defined by this 4-vector and the photon 4-momentum $p^\mu$ at the lamp-post location. Hence, in order to find the polarization angle for each distant observer or at each point of the disc, i.e. to find three components of the polarization 4-vector $f^\mu$ in the local reference frame, these three unknown components can be expressed {\it at the receiving point} by means of the $\kappa_1$ and $\kappa_2$ quantities, while $\kappa_1$ and $\kappa_2$ can be expressed by means of the initial $p^\mu$ and $f_0^\mu$ of each photon {\it at the emission point} due to the conservation of $K_\textrm{S}$. Last but not least, we have to construct the local polarization angle $\chi$ according to its definition at the receiving point.

We will derive the final expressions in the Kerr space-time with metric coefficients $g_{\mu \nu}$, using the global Boyer-Lindquist coordinates $\{ t, r, \theta, \varphi   \}$, see \cite{Boyer1967}\footnote{Note that the inclination $i$ of the disc with respect to a distant observer coincides with the coordinate $\theta$ of the arriving photon.}. We track only the photons from the rotation axis, thus the Carter's constants of motion \citep{Carter1968} are in a reduced form $l = \alpha \sin{i}= 0$, $q = \sqrt{\beta^2 + (\alpha^2-a^2)\cos^2{\!i}} = \sqrt{\beta^2 - a^2\cos^2{\!i}}$ with $\alpha$ and $\beta$ being the impact parameters measured perpendicular to and parallel with the spin axis of the black hole projected onto the observer’s sky \citep{Cunningham1973}, respectively. We follow \cite{Carter1968, Misner1973} for the expressions of $p_\mu$ components. While $p_\varphi = 0$ is trivial and we normalize the conserved energy as $p_t = -1$, the other photon momentum components are $|p_\theta|=\sqrt{q^2+a^2\cos^2\!\theta}$ and $|p_r|=\sqrt{(r^2+a^2)^2-\Delta(q^2+a^2)}/\Delta$ with $\Delta=r^2-2r+a^2$. Note that to connect the final photon destination point at the observer (defined by the observer inclination, $i$) or at the disc (defined by the radius, $r$) with the initial photon momentum at the primary source (defined by the height, $h$, and constant of motion, $q$), we still need to perform ray-tracing from the lamp-post to the observer and to the disc.

\begin{figure}
	\includegraphics[width=1.\columnwidth]{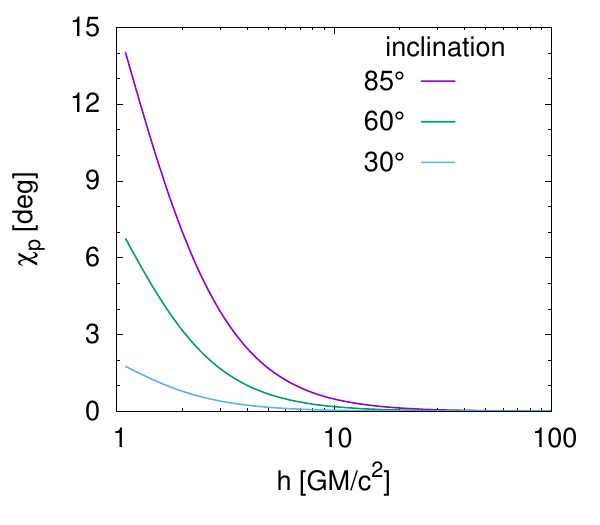}
	\caption{Change of the polarization angle, $\chi_{\rm p}$, between the lamp-post and the observer versus lamp-post height $h$ for BH spin $a = 1$ and three different observer's inclinations $i = 30^{\circ},\,60^{\circ}$, and $85^{\circ}$.}
	\label{fig:chiao}
\end{figure}

\medskip
Having laid down the procedure, let us first solve the case of transfer from the lamp to a distant observer and derive expressions for the change of polarization angle $\chi_\textrm{p}$ at spatial infinity with respect to the initial polarization angle $\chi_0 = 0$ at the lamp location, i.e. initial polarisation vector parallel to the projection of symmetry axis into the polarization plane. We may express $\kappa_1$ and $\kappa_2$ on the rotation axis of symmetry (i.e. at the lamp-post) as
\begin{equation}\label{k10}
    \kappa_1 = h\sqrt{\frac{q^2+a^2}{h^2+a^2}},
\end{equation}
\begin{equation}\label{k20}
    \kappa_2 = -a\sqrt{\frac{q^2+a^2}{h^2+a^2}},
\end{equation}
where we set $f_{0\,t} = f_{0\,\varphi} = 0$ and the other two components, $f_{0\,r}$ and $f_{0\,\theta}$ are expressed from general conditions of (i) polarisation vector being perpendicular to the momentum, $f^\mu p_\mu = 0$, and (ii) polarisation vector being normalised to unity, $f_\mu f^\mu = 1$. Now let us perform the transformation (\ref{reduction}) at spatial infinity (using the conditions $f^\mu p_\mu = p^\mu p_\mu = 0$)
\begin{equation}\label{trafo1}
    f'^{\textrm{(t)}} = 0,
\end{equation}
\vspace*{-6mm}
\begin{equation}
    f'^{\textrm{(r)}} = r^2p^\theta (p^\theta f^r - p^r f^\theta),
\end{equation}
\vspace*{-6mm}
\begin{equation}
    f'^{(\theta)} = rp^r (p^\theta f^r - p^r f^\theta ),
\end{equation}
\vspace*{-6mm}
\begin{equation}\label{trafo4}
    f'^{(\varphi)} = r\sin{i}f^{\varphi},
\end{equation}
where we transformed $f^\mu$ and $p^\mu$ into the local reference frame defined by the orthonormal tetrade $e^{\textrm{(t)}}_\mu \equiv (-1,0,0,0)$, $e^{\textrm{(r)}}_\mu \equiv (0,1,0,0)$, $e^{(\theta)}_\mu \equiv -r\,(0,0,1,0)$, $e^{(\varphi)}_\mu \equiv r\sin{i}\,(0,0,0,1)$. Combining (\ref{trafo1})--(\ref{trafo4}) with (\ref{KS}) yields $\kappa_1$ and $\kappa_2$ for each geodesic at spatial infinity (i.e. at the observer's location)
\begin{equation}\label{k1}
    \kappa_1 = \beta f'^{(\theta)} + a\sin{i}f'^{(\varphi)},
\end{equation}
\vspace*{-6mm}
\begin{equation}\label{k2}
    \kappa_2 = -a\sin{i}f'^{(\theta)}+\beta f'^{(\varphi)}.
\end{equation}
Putting (\ref{k1}) equal to (\ref{k10}) and (\ref{k2}) equal to (\ref{k20}) brings the expression for the change of the polarization angle $\chi_{\rm p}$ at the local reference frame of the observer
\begin{equation}
    \tan{\chi_{\rm p}} \equiv -\frac{f'^{(\varphi)}}{f'^{(\theta)}} = a\frac{\beta - h\sin{i}}{a^2\sin{i}+h\beta}.
\end{equation}
The dependence of $\chi_{\rm p}$ on height is visualized in Figure \ref{fig:chiao} for maximally spinning BH and three observer inclinations.

\medskip
Now let us focus on the transfer from the lamp to the disc. We evaluate the $\kappa_1$ and $\kappa_2$ constants from the definition of $K_{\rm S}$ by (\ref{KS}) in the equatorial plane (i.e. at the disc)
\begin{equation}\label{kd1}
    \kappa_1 = \frac{r^2+a^2}{r}p_r f_t + \frac{r^2+a^2}{r} f_r + \frac{ap_r}{r}f_\varphi,
\end{equation}
\vspace*{-3mm}
\begin{equation}\label{kd2}
    \kappa_2 = \frac{ap_\theta}{r}f_t + \frac{a}{r}f_\theta + \frac{p_\theta}{r}f_\varphi.
\end{equation}
Along with the conditions $p_\mu p^\mu = 0$, $f^\mu p_\mu = 0$ and $f_\mu f^\mu = 1$, we may express $f_t$, $f_\theta$, $f_\varphi$ as functions of $\kappa_1$, $\kappa_2$, $p^\theta$, $p^r$ and the metric coefficients. This approach turns out to simplify the computations. Instead of performing transformation (\ref{reduction}), we will subtract the $r$-component of $f_\mu$ in the equatorial plane
\begin{equation}
    f'_\mu = f_\mu - \frac{f_r}{p_r}p_\mu,
\end{equation}
because this transformation largely simplifies expressions. Since the polarization vector is perpendicular to the photon momentum and definition of polarization angle is in the plane perpendicular to the photon momentum, any transformation $f'_\mu = f_\mu + A p_\mu$ does not change its value (the tetrade vectors $e_{\rm (y)}^\mu$ and $e_{\rm (z)}^\mu$, defined below, into which we project the polarization vector $f'^\mu$ in the angle definition (\ref{eq:angle}) are perpendicular to the photon momentum). This yields the reduced components of $f'_\mu$ in the form of
\begin{equation}\label{dtrafo1}
    f'_t = -\frac{\Delta p^\theta}{ar(p^r)^2}\kappa_2 + \left[\frac{(r^2+a^2)}{a}\frac{\Delta (p^\theta)^2}{(p^r)^2} + g_{t\varphi}\right]\frac{f_\varphi}{g_{\varphi \varphi}},
\end{equation}
\vspace*{-3mm}
\begin{equation}\label{dtrafo2}
    f'_{r} = 0,
\end{equation}
\vspace*{-5mm}
\begin{equation}\label{dtrafo3}
    f'_{\theta} = \frac{g_{\varphi \varphi}}{ar(p^r)^2}\kappa_2 - \frac{(r^2+a^2)p^\theta}{a(p^r)^2}f_\varphi,
\end{equation}
\vspace*{-2mm}
\begin{equation}\label{dtrafo4}
    f'_{\varphi} = f_\varphi = \frac{arp^r \kappa_1 + r(r^2+a^2)p^\theta \kappa_2}{q^2+a^2},
\end{equation}
where $g_{t\varphi} = -2a/r$, $g_{\varphi \varphi} = [(r^2+a^2)^2-\Delta a^2]/r^2$ are taken from the Kerr metric in the equatorial plane. We assume Keplerian accretion disc \citep{Novikov1973}, where the components of the 4-velocity are $U^t=(r^2+a\sqrt{r})/r/\sqrt{r^2-3r+2a\sqrt{r}}$, $U^r=U^\theta=0$ and $U^\varphi=\Omega_{\rm K} U^t$ with relativistic Keplerian rotational velocity $\Omega_{\rm K}=(r^{3/2}+a)^{-1}$. 
To define the polarization angle $\chi_\textrm{d}$, we need to construct an orthonormal tetrade in the local co-moving frame with the disc. We choose the time tetrade vector to be defined by the 4-velocity of the orbiting material, the first spatial tetrade vector to be directed in the direction of motion of the incoming photon, the second spatial tetrade vector to be defined by the projection of the local disc normal into the plane perpendicular to the photon momentum, and the last spatial tetrade vector is created as the vector product of the others so that the full orthonormal tetrade is formed. 
\begin{figure}
	\includegraphics[width=1.\columnwidth]{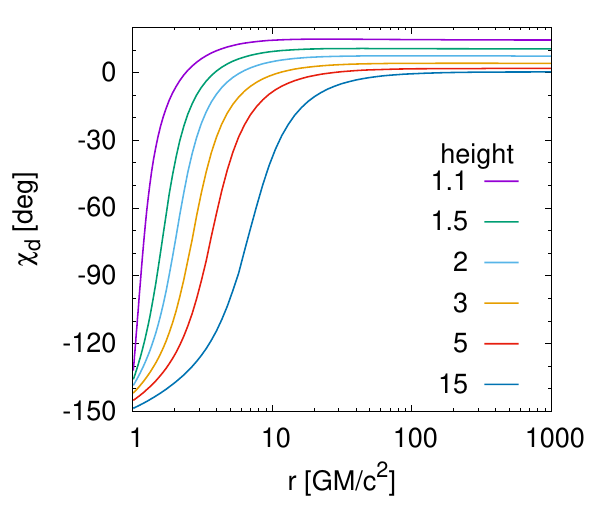}
	\caption{Change of the polarization degree $\chi_{\rm d}$ between the lamp-post and the disc as a function of radius $r$ of the disc in the equatorial plane for BH spin $a = 1$ and for several different lamp-post heights.}
	\label{fig:chiad}
\end{figure}
Thus we set
\begin{equation}\label{de1}
    e_{\rm (t)}^\mu \equiv U^\mu,
\end{equation}
\vspace*{-5mm}
\begin{equation}
    e_{\rm (x)}^\mu \equiv \frac{p^\mu+p^\nu U_\nu U^\mu}{-p_\sigma U^\sigma},
\end{equation}
\vspace*{-2mm}
\begin{equation}\label{de3}
    e_{\rm (y)}^\mu \equiv \frac{n^\mu - \frac{n_\nu p^\nu}{-p_\sigma U^\sigma}e_{\rm (x)}^\mu}{\sqrt{1-\left(\frac{n_\rho p^\rho}{p_\iota U^\iota}\right)^2}},
\end{equation}
\vspace*{-2mm}
\begin{equation}\label{de4}
    e_{\rm (z)}^\mu \equiv \frac{\varepsilon^{\alpha \beta \gamma \mu} U_\alpha p_\beta n_\gamma}{-p_\sigma U^\sigma\sqrt{1-\left(\frac{n_\rho p^\rho}{p_\iota U^\iota}\right)^2}},
\end{equation}
where $\varepsilon^{\alpha \beta \gamma \mu} = -\delta^{\alpha \beta \gamma \mu}/\sqrt{-g}$ is the permutation tensor defined by the permutation symbol, $\delta^{\alpha \beta \gamma \mu}$, and the metric tensor determinant, $\sqrt{-g} = r^2$, and $n^\mu = (0,0,-1/r,0)$ is the local disc's normal. Then using the relations $f^\mu p_\mu = 0$, $U_\theta = 0$, $p_\varphi = 0$, (\ref{dtrafo2}), (\ref{de1})--(\ref{de4}) we arrive at the expression for the change of polarization angle $\chi_\textrm{d}$ at the local reference frame of the observer co-moving with the disc with respect to the initial polarization angle $\chi_0 = 0$ at the lamp location
\begin{equation}
\label{eq:angle}
    \tan{\chi_\textrm{d}} \equiv \frac{e_{\rm (z)}^\mu f'_\mu}{e_{\rm (y)}^\mu f'_\mu} = \frac{U_t p_r f'_\varphi + U_r f'_\varphi - U_\varphi p_r f'_t}{p_\nu U^\nu f'_\theta-q U^\mu f'_\mu},
\end{equation}
where $U^\mu f'_\mu = U^t f'_t + U^\varphi f'_\varphi$, 
$p_\nu U^\nu = -U^t$, 
$f'_t$, $f'_\theta$ and $f'_\varphi$ follow from (\ref{dtrafo1}), (\ref{dtrafo3}) and (\ref{dtrafo4}), respectively, where the constants $\kappa_1$ and $\kappa_2$ have to be inserted from (\ref{k10}) and (\ref{k20}), respectively. The radial dependence of $\chi_{\rm d}$ is visualized in Figure \ref{fig:chiad} for maximally spinning BH and several lamp-post heights.

\section{Correction of local disc reflection computations}\label{loc_error}

\begin{figure}
\includegraphics[width=\columnwidth]{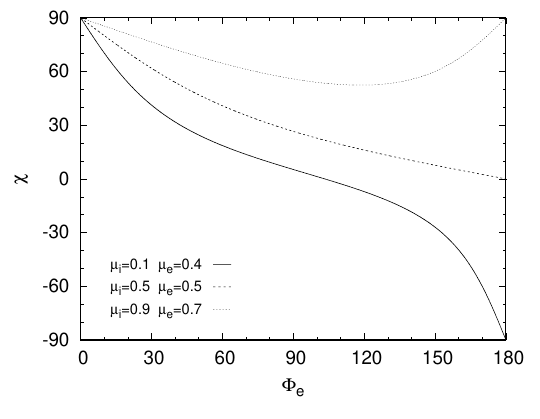}
\caption{Correction of figure 4 (right panel) in \citet{Dovciak2011}. The emergent polarization angle $\chi$ in the local co-moving frame and its dependence on the emergent azimuthal angle $\Phi_\textrm{e}$ for three pairs of incident and emergent cosines of inclination angles $\mu_\textrm{i}$ and $\mu_\textrm{e}$.}
\label{fig:local_pol_angle_remake}
\end{figure}

We note that a sign error was present in the definition of local polarization angle in polarization induced by reflection computations in \cite{Dovciak2011}, using the Chandrasekhar's formulae \citep{Chandrasekhar1960}. This correction does not propagate to the main results of the current paper, but it should be borne in mind, if one attempts to compare the latest results to the ones in \cite{Dovciak2011}. The Stokes parameter $-U$ was assumed instead of $U$ for the emergent light rays in the local co-moving frame. Thus, right panel of figure 4 in \cite{Dovciak2011} should have an opposite $y$-axis, and we correctly display it here in Figure \ref{fig:local_pol_angle_remake}. This local error then propagated into the global computations of both polarization angle and degree, i.e. figures 6 and 8--14 in \cite{Dovciak2011}.

\bsp	
\label{lastpage}
\end{document}